\documentclass[letterpaper,twocolumn,10pt]{article}
\usepackage{usenix2019_v3}

\usepackage{tikz}
\usepackage{amsmath}
\usepackage[small,compact]{titlesec}

\usepackage{filecontents}
\usepackage{xspace}
\usepackage{booktabs} %
\usepackage{makecell}
\usepackage{url}
\usepackage{subfig}
\usepackage{ulem} %
\usepackage[export]{adjustbox} %
\usepackage{multirow}
\usepackage{enumitem}
\usepackage{placeins}

\newif\ifdraft
\drafttrue %

\renewcommand{\emph}[1]{\textit{#1}}

\newcommand{\pllm}{NanoFlow\xspace}
\newcommand{\newparall}{intra-device parallelism\xspace}

\newcommand{\NewParall}{Intra-device Parallelism\xspace}

\newcommand{\oldparall}{inter-device parallelism\xspace}

\newcommand{\lm}{LLaMA}

\newcommand{\nano}{nano-batch}
\newcommand{\Nano}{Nano-batch}
\newcommand{\nanoop}{nano-operation}

\newcommand{\auto}{auto-search}
\newcommand{\Auto}{Auto-search}

\newcommand{\fig}[1]{Figure~\ref{#1}}

\newcommand{\kv}{KV-cache}

\newcommand{\outputlen}{output length in tokens}

\newcommand{\blue}[1]{\textcolor{black}{#1}}
\newcommand{\green}[1]{\textcolor{black}{#1}}

\newcommand{\red}[1]{\textcolor{red}{#1}}

\newcommand{\computeoptimalratio}{$68.5$\%\xspace}
\newcommand{\computesota}{$1.91\times$\xspace}
\newcommand{\requestsota}{$1.64\times$\xspace}
\newcommand{\benefitnetwork}{$1.07\times$\xspace}
\newcommand{\benefittotal}{$1.17\times$\xspace}

\newcommand{\nanooverhead}{$13.2$\%\xspace}
\newcommand{\portmin}{$50$\%\xspace}
\newcommand{\portmax}{$72$\%\xspace}
\newcommand{\portavggain}{$2.66\times$\xspace}

\begin{document}

\date{}

\title{\Large \bf \pllm{}: Towards Optimal Large Language Model Serving Throughput}
\author{
{\rm Your N.\ Here}\\
Your Institution
\and
{\rm Second Name}\\
Second Institution
} %

\author{
{\rm Kan Zhu}\\
University of Washington
\and
{\rm Yufei Gao}\\
University of Washington \\ Tsinghua University
\and
{\rm Yilong Zhao}\\
University of Washington \\ UC Berkeley
\and
{\rm Liangyu Zhao}\\
University of Washington
\and
{\rm Gefei Zuo}\\
University of Michigan
\and
{\rm Yile Gu}\\
University of Washington
\and
{\rm Dedong Xie}\\
University of Washington
\and
{\rm Tian Tang}\\
University of Washington \\ Tsinghua University
\and
{\rm Qinyu Xu}\\
University of Washington \\ Tsinghua University
\and
{\rm Zihao Ye}\\
University of Washington
\and
{\rm Keisuke Kamahori}\\
University of Washington
\and
{\rm Chien-Yu Lin}\\
University of Washington
\and
{\rm Ziren Wang}\\
University of Washington \\ Tsinghua University
\and
{\rm Stephanie Wang}\\
University of Washington
\and
{\rm Arvind Krishnamurthy}\\
University of Washington
\and
{\rm Baris Kasikci}\\
University of Washington
}

\maketitle

\begin{abstract}
Large Language Models (LLMs) have resulted in a surging demand for planet-scale serving systems, where tens of thousands of GPUs continuously serve hundreds of millions of users. Consequently, throughput has emerged as a key metric that determines serving systems' performance. Due to large model sizes and memory-intensive self-attention, LLM serving has been commonly assumed to be memory-bound. Through a detailed analysis, we show that despite having memory-intensive components, end-to-end LLM serving is compute bound for most common workloads and LLMs. Alas, most existing serving engines fall short from optimal compute utilization, because the heterogeneous operations that comprise LLM serving---compute, memory, networking---are executed sequentially within a device.

We propose \pllm{}, a novel serving framework that exploits \newparall{}, which overlaps the usage of heterogeneous resources within a single device.
\pllm{} splits inputs into smaller \textit{\nano{}es} and duplicates operations to operate on each portion independently, enabling overlapping. \pllm{} automatically identifies the number, size, ordering, and GPU resource allocation of \nano{}es to minimize the execution time, while considering the interference of concurrent operations. 
We evaluate \pllm{}'s end-to-end serving throughput on several popular models such as \lm{}-2-70B, Mixtral 8$\times$7B, \lm{}-3-8B, etc. With practical workloads, \pllm{} provides \computesota{} throughput boost compared to state-of-the-art serving systems, achieving between \portmin to \portmax of optimal throughput across popular models.

\end{abstract}

\section{Introduction}
\label{sec:intro}

Large language models (LLMs) based on transformers power applications like chatbots, search engines, and office software but are highly resource-intensive~\cite{Mehdi_2023, Spataro_2023, ChatGPT}. Reports indicate over $200$ million weekly ChatGPT users, with API usage doubling after GPT-4o Mini's release~\cite{seminanalysis,demandsageChatGPTStatistics,openai_chatgpt_verge}. Given the massive demand and limited GPU availability~\cite{gpushortage, nytgpushortage}, maximizing hardware utilization is critical. In particular, throughput, measured as tokens per device per second, has become a key factor in reducing serving costs~\cite{vllm, deepspeed-fastgen, tensorrtllm}.

LLMs are more resource-intensive than earlier DNNs due to unique characteristics. \green{Their model sizes are vastly larger, such as GPT-3 with 175B parameters requiring $5\times$A100 80 GB GPUs to store weights in 16-bit precision, with some closed-source models even larger~\cite{brown2020language}.} \green{Moreover, LLMs' self-attention mechanism~\cite{attention} scales total memory loading quadratically with the context length---the aggregate length of input and output sequences---during token generation.} LLM serving systems often cache per-request state in a \emph{key-value cache} (\kv{}), whose size can surpass the size of model weights~\cite{tang2024quest}, driving up memory demands.
Moreover, each LLM iteration loads the entire model weights plus the \kv{}, while only producing one output token for each decoding sequence, further increasing memory pressure.
Due to these characteristics, \blue{LLM serving is commonly assumed to be overall memory bandwidth bound.}

In this paper, we show that while the self-attention operation is indeed memory-bound, \emph{LLM serving is often compute-bound when considered as a whole}.
We provide a detailed analysis (which we also validate empirically) of the common operations used in current LLMs and find that:
(1) Batching prefill and decode requests amortizes weight loading, making general matrix multiplications (GEMMs) compute-bound.(2) Although batched decode request still need to load a unique \kv{} per sequence, novel optimizations such as grouped query attention~\cite{ainslie2023gqa} (GQA) can reduce the memory loads. (3) As model sizes grow, the compute operations from GEMMs tend to dominate compared to self-attention. 
Consequently, for many common workloads 
and LLMs, the total compute operations dominate network and memory I/O.

However, in practice, we find that LLM serving engines are far from optimal in terms of compute utilization when measured end-to-end and compared to the hardware FLOPs.
This is because LLMs use \emph{heterogeneous operations}.
For example, the self-attention is memory-bound during the decode phase.
\blue{Also, when models are larger than a single GPU's memory capacity, GEMM is often split across multiple GPUs when using tensor parallelism, and thus requires network communication in addition to compute.}
We find that in current LLM serving engines, individual operations have high utilization of their bottlenecked resource ($\sim80\%$).
However, because each operation is executed one at a time, and different operations have different resource bottlenecks, the total compute utilization is only 40\%, as shown in Section \ref{sec:eval-offline}.

To address this gap, we present \pllm{}, a serving framework that aims to maximize utilization of the workload's resource bottleneck, when considered as a whole.
The key insight is to leverage \emph{intra-device parallelism for heterogeneous operations}.
In particular, \pllm{} splits each input batch into \textit{\nano{}es} and duplicates operations across \nanoop{}s, with each \nanoop{} processing a single \nano{}. Since \nanoop{}s operate on separate \nano{}es without dependencies, heterogeneous operations---such as memory-bound and compute-bound operations---can execute \emph{simultaneously} rather than sequentially. This approach facilitates fine-grained pipelining across compute, memory, and network resources within a \emph{single device}. As \nanoop{}s duplicate each operation, this method increases memory I/O for weight loading.
However, when the workload is compute-bound as a whole, the additional memory I/O can be hidden via pipelining.

To achieve efficient pipelining, \pllm{} executes \nanoop{}s with a fraction of the GPU's resources.
For instance, while most of the GPU resources can be assigned to a compute-heavy operation for high compute utilization, a memory-bound operation can still achieve considerable memory bandwidth usage using the remaining GPU resources.

Designing \pllm{} presents several major challenges. First, given the multitude and diversity of LLMs, determining the appropriate number of \nano{}es, the input batch size per each \nanoop{}, and the execution order of \nanoop{}s involves navigating a large search space. Second, when memory- or network-bound kernels run alongside compute-bound kernels, competition for GPU resources (e.g., execution units, caches) causes unpredictable performance interference, slowing down kernels running in parallel~\cite{orion}. Balancing these trade-offs is critical. Allocating more resources to compute-bound kernels may excessively delay memory- or network-bound kernels, while prioritizing memory- or network-bound kernels could hinder overall throughput by slowing down compute-bound operations.

To tackle these challenges and efficiently explore the large search space, \pllm{} proposes \textit{auto-search} to automatically construct an intra-device pipeline. %
The \auto{} process reduces the search space by executing in two stages, which approximates the optimal solution balancing efficiency and accuracy.
In the first stage, \pllm{} determines the initial pipeline schedule, including the number, size, and ordering of \nanoop{}s, assuming no interference between kernels.
In the second stage, \pllm{} refines the pipeline by profiling actual kernel interference and re-planning accordingly. \pllm{} also provides a runtime system to execute the optimized pipeline by efficiently forming \nano{}es, allocating GPU resources, and managing the \kv{}.

We provide a detailed evaluation of \pllm{} on the \lm{}-2-70B model~\cite{llama2} using one NVIDIA $8\times$A100 DGX node~\cite{dgx}. For practical workloads
including ShareGPT~\cite{sharegpt}, LMSys~\cite{zheng2023lmsyschat1m} and Splitwise~\cite{splitwise}, \pllm{} achieves on average \computesota{} greater throughput compared with state-of-the-art serving frameworks including vLLM \cite{vllm}, DeepSpeed-FastGen \cite{deepspeed-fastgen} and TensorRT-LLM \cite{tensorrtllm}, which is \computeoptimalratio{} of the theoretically optimal throughput we derive in \S\ref{sec:analysis}. \pllm{} achieves similar latency compared with the best baseline TensorRT-LLM at a low request rate, while handling up to \requestsota{} higher request rates within SLO constraint. We also demonstrate \pllm{}'s ability to automatically generate efficient pipelines for other models and architectures such as Mixture-of-Experts (MoE), by evaluating popular LLMs (\lm{}-3-70B, \lm{}-3-8B, QWen2-72B, Deepseek-67B, and Mixtral 8$\times$7B). \pllm{}-generated serving pipelines achieve \portmin-\portmax of optimal throughput, which on average reaches \portavggain throughput gain compared with vLLM.

\begin{figure*}[t]
    \centering
    \includegraphics[width=\linewidth]{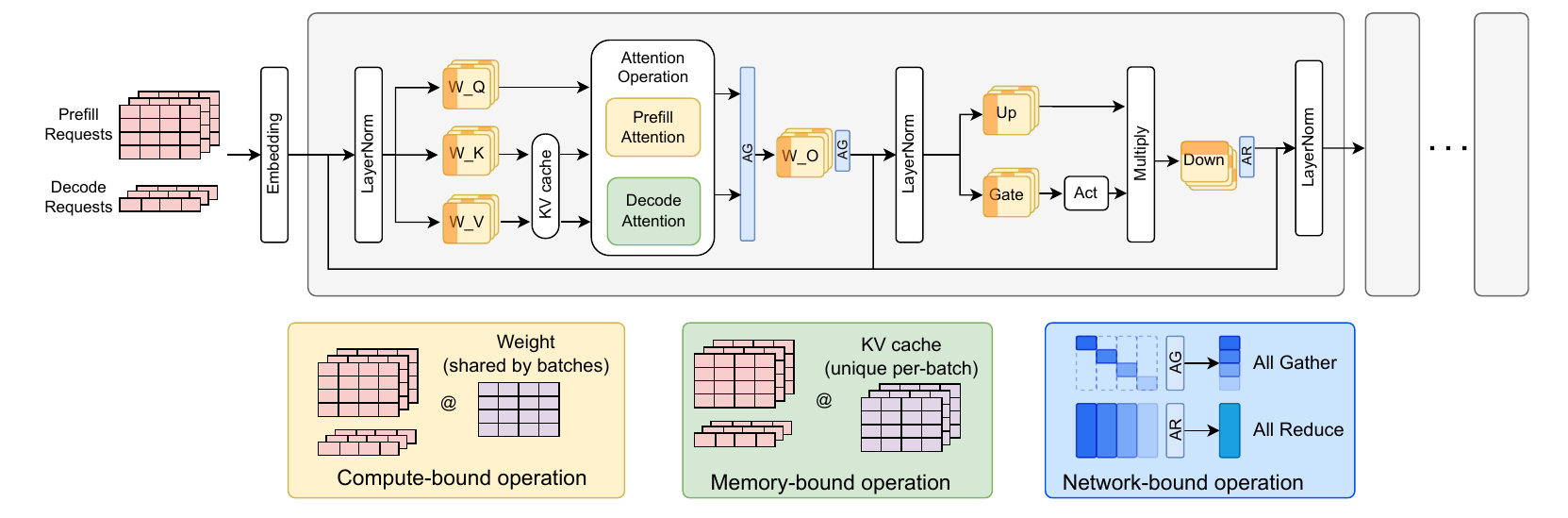}
    \caption{Transformer architecture. The operations in the yellow boxes have large batch sizes and share model weight parameters across requests; hence, they are compute-bound. Operations in green boxes require loading a unique KV cache for each request; hence, they are memory-bound. The blue box represents network operations that perform synchronization between operations.}
    \label{fig:workflow}
\end{figure*}
In summary, we contribute the following:
\begin{itemize}[noitemsep, topsep=0pt]
    \item A detailed analysis and empirical validation of the workload characteristics and theoretically optimal throughput of LLM serving systems that shows modern LLMs operate in a compute-bound regime. 
    \item \pllm{}, that comprises (1) an auto-search engine that automatically computes a pipeline for \nano{}es to improve LLM serving throughput, (2) an end-to-end LLM serving runtime.
    \item A comprehensive evaluation of \pllm{} that achieves \computesota{} throughput gain compared to baselines and \computeoptimalratio{} of theoretical maximum throughput.
\end{itemize}

\section{Background}

\subsection{LLM Inference Workflow}
\label{sec:bg:inference}

Recent LLMs like GPT-4, \lm{}, and Mistral~\cite{gpt4,llama2,mistral} are based on the decoder-only transformer~\cite{attention}. The inference process of these LLMs has two phases~\cite{orca}: (1) \textit{prefill}, which processes the input prompt all at once, and (2) \textit{decode}, which generates output tokens one at a time auto-regressively. The prefill phase initializes the \textit{\kv{}}~\cite{kvcache}, which keeps the per-request state to speed up the decode phase. 

Both the prefill and decode phases utilize the same inference workflow in \fig{fig:workflow}. Upon receiving an inference request, the input traverses identical \textit{decoder layers} to produce the subsequent tokens. In each decoder layer, during the attention stage, the input is multiplied by weight matrices $W_Q$, $W_K$, and $W_V$ to form Query, Key, and Value, with Key and Value concatenated into the existing \kv{}. The Query is then multiplied with the existing Keys and normalized to assess the similarity between the current token and all previous tokens. This similarity is employed to compute the weighted average of the Value, aggregating the context information. The outcome undergoes O-projection through a linear transformation using $W_O$ followed by a layer normalization operation~\cite{layernorm}. Subsequently, during the feed forward network stage, the activation is multiplied by $W_{up}$ and $W_{gate}$ separately to generate $o_{u}$ and $o_{g}$. Next, $o_{g}$ passes through an activation function (e.g., SiLU~\cite{silu}), followed by an element-wise multiplication with $o_{u}$. Finally, $W_{down}$ is applied as Down-projection to generate the output, which serves as the input to the next layer.

\subsection{Operation Characteristics}
According to the characteristics of the operations during inference, we can classify them into four categories.

\textbf{Dense operations.} 
The KQV generation, O projection, Up, Gate, and Down projection compute the multiplication of activation and model weights, which we define as \textit{dense operations}. In the prefill phase, all tokens of new requests form a large batch for these dense operations. In the decode phase, batched decoding aggregates the activations from all the decode requests for dense operations~\cite{orca}. Moreover, as the prefill phase and decode phase share weight matrices, the activations from both phases can be combined to further exploit the batching effect by amortizing the weight loading cost~\cite{sarathi}. As a result, dense operations are \textit{compute-bound}.

\textbf{Attention operations.} Transformers capture token relations through attention~\cite{attention}. The prefill phase processes all input tokens simultaneously, making it \textit{compute-bound}, while the decode phase generates one token per iteration, loading cached $K$ and $V$ vectors from the \kv{}, making it \textit{memory-bound}.

\textbf{Network operations.}
Collective communication operations including AllGather (AG) and AllReduce (AR) are \textit{network-bound} and call for high bandwidth interconnection between nodes like NVLink~\cite{nvlink}.

\textbf{Other operations.}
We classify the rest of the operations in transformer architecture, e.g., layer norms, position embeddings, etc., as other operations, due to their short runtime compared to dense or attention operations. 

\subsection{Serving LLMs at Scale}
\label{backgroud:serving}

With large models, multiple GPUs are required to provide enough resources, including both memory and compute, for efficient LLM serving~\cite{alpaserve}. Existing works~\cite{alpa,shoeybi2020megatronlm,he2022fastermoe} have comprehensive evaluations of different \textit{\oldparall{}} paradigms:

\textit{Tensor Parallelism}~\cite{shoeybi2020megatronlm,alpa} splits each weight matrix onto different GPUs and executes each operation collectively, avoiding weight duplication, thereby scaling compute capacity. However, frequent collective communications after operations are required to synchronize each operation's results, as shown in \fig{fig:workflow}.

\textit{Pipeline Parallelism}~\cite{huang2019gpipe}
splits the model into stages by layers so that each GPU only needs to hold part of the weights. While all devices with tensor parallelism execute on the same batch of data, devices with pipeline parallelism operate on different micro-batches of data split by stages.

In practice, these parallelism paradigms are used in combination. Tensor parallelism is widely used on GPUs within a node to enable a larger batch size, while pipeline parallelism is used across nodes to further scale the cluster.

\section{Analysis}
\label{sec:analysis}

We now introduce the factors determining the throughput of LLM serving (\S \ref{sec:costfactors}) and model the cost of LLM serving (\S \ref{sec:costmodel}). We then classify modern LLMs serving workloads according to our model (\S\ref{sec:classification}) and empirically validate this classification (\S\ref{sec:validation}), showing that popular workloads are compute-bound. Then, for the compute-bound scenarios, we determine optimal serving throughput (\S \ref{sec:analysis-optimal}), explain why existing systems fall short of the optimal (\S \ref{sec:analysis-existing}), and how this motivates our proposed approach, \newparall{} (\S \ref{sec:analysis-intra}).

\subsection{Key Factors of Serving Throughput}
\label{sec:costfactors}

We define throughput as the \emph{total throughput} of serving, which is the number of tokens processed per second by both prefill and decode phases. Throughput is affected by hardware properties, model configuration, user queries, and batch size.

\textbf{Hardware specification.}
The following factors determine the available GPU hardware resources:
\begin{itemize}[noitemsep, topsep=0pt]
    \item $N_{GPU}$: number of GPUs
    \item $MemBW$ (GB/s): aggregate GPU memory bandwidth
    \item $MemSize$ (GB): aggregate GPU memory capacity
    \item $Compute$ (GFLOP/s): aggregate GPU compute capacity
    \item $NetBW$ (GB/s): aggregate GPU interconnect bandwidth
\end{itemize}

\textbf{Model configuration. } The following factors regarding the model architecture determine the computation, memory, and network demand when serving LLMs:
\begin{itemize}[noitemsep, topsep=0pt]
    \item $D_{model}$: hidden dimension size
    \item $L$: number of layers
    \item $P_{Model}$: number of parameters
    \item $R_{GQA}$: group size of GQA
    \footnote{The number of attention heads with a shared KV head~\cite{ainslie2023gqa}.}
    \item $S_{type}$ (Bytes): number of bytes (size) of the data type for model parameters (e.g., $2$ for FP16)
\end{itemize}

\textbf{User query statistics.} The following user input statistics also affect the system throughput:
\begin{itemize}[noitemsep, topsep=0pt]
    \item $p$: average number of tokens in prompts to be prefilled
    \item $d$: average number of tokens in output to be decoded
\end{itemize}

Therefore, a serving request from a user corresponds to $p$ prefill tokens and $d$ decode tokens on average, for a total of $p+d$ tokens that we use when calculating the total throughput. Using user query statistics, total throughput (tokens generated per second) can be easily converted to other throughput metrics, for example,  decoding throughput (decode output per second) is $\frac{d}{p+d} \times \mathrm{Throughput_{total}}$, and request per second is $\frac{1}{p + d}\times\mathrm{Throughput_{total}}$.

\textbf{Batch size.} An optimal throughput-oriented serving system needs to process large batches. For compute-bound dense operations, e.g., GEMMs, a larger batch amortizes the weight loading overhead and increases execution unit occupancy, which provides higher compute utilization. For memory- and network-bound operations, all requests share the same warmup cost (e.g., kernel launching overhead, pipeline setup overhead, network synchronization overhead, etc.), so a larger batch size leads to higher throughput. Therefore, when considering the optimal system throughput, we assume that the system always operates at the largest batch size at which the total available memory can hold the model weights and all the KV caches for the requests. \footnote{Activations occupy negligible (less than 5\%) of the memory\cite{vllm}, thus we omit them for simplicity.}

\subsection{Cost Model of LLM Serving}
\label{sec:costmodel}

Under the assumption of maximum batch size, we derive the latency of an LLM serving iteration (where a batch of user requests are being processed by all the transformer layers) from the perspectives of required memory, compute, and network resources.

\noindent \textbf{Memory.} From the memory perspective, latency is:
\begin{align}
    \label{eq:mem-latency}
    T_{mem} = \frac{MemSize}{MemBW}
\end{align}

This is because the entire device memory content needs to be loaded into the GPU caches and registers once
for each iteration (typical in modern serving systems~\cite{vllm}) when using the largest batch size. Although the model weights are shared between iterations, due to the long reuse distance, it is infeasible to cache the model weights and avoid loading from the device memory.

\noindent \textbf{Compute.} 
In LLM serving, all operations contribute to the total compute, while Dense operations produce the vast majority of computations (as we validate in \S\ref{sec:validation}). Therefore we model the compute latency of dense operations below\footnote{Our supplementary material provides a more detailed analytical model.}.

\begin{figure*}[t]
    \centering
    \includegraphics[width=\textwidth]{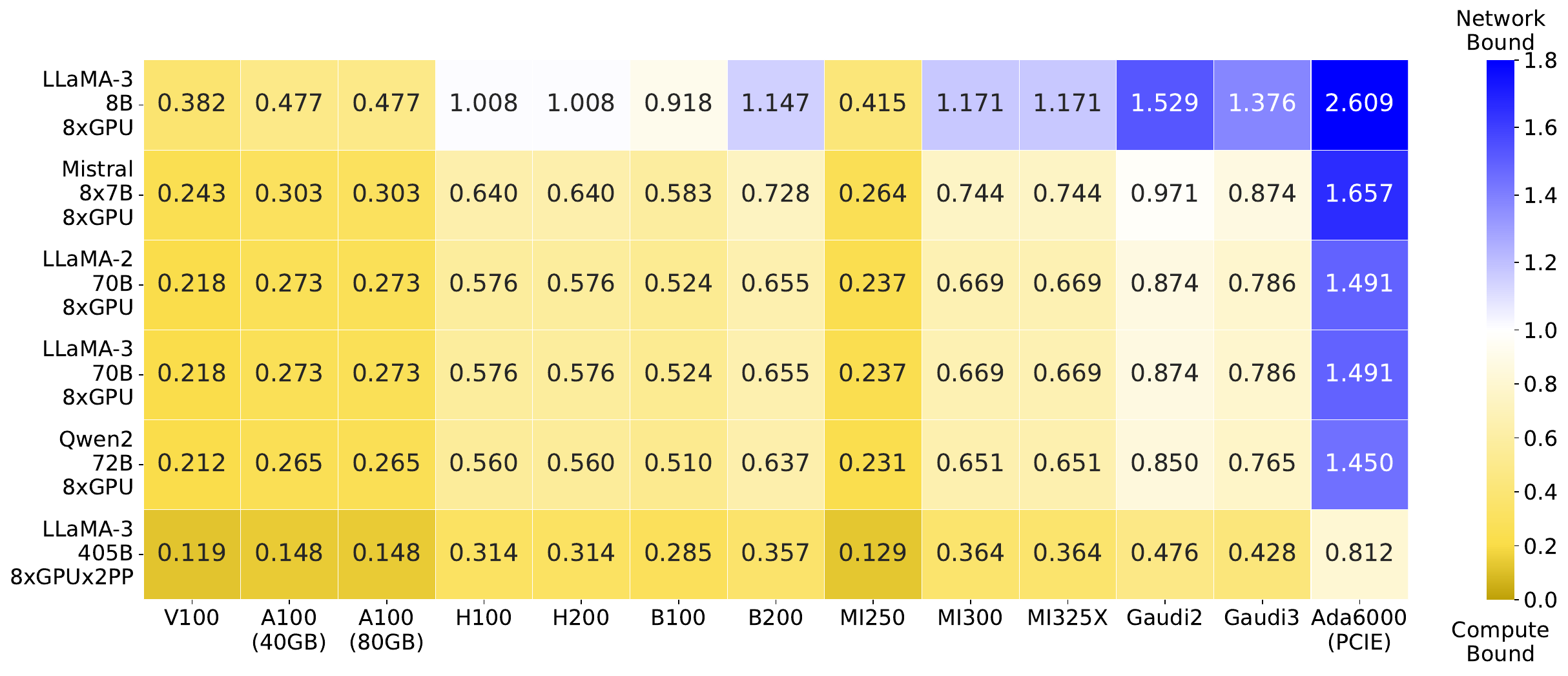}
    
    \caption{\blue{Comparison of network time and compute time. The closer to yellow, the more compute-bound the workload is, whereas the closer to blue indicates the workload is more network-bound.}}
    \label{fig:net-compute}
\end{figure*}

\begin{figure}[t]
    \centering
    \includegraphics[width=\columnwidth]{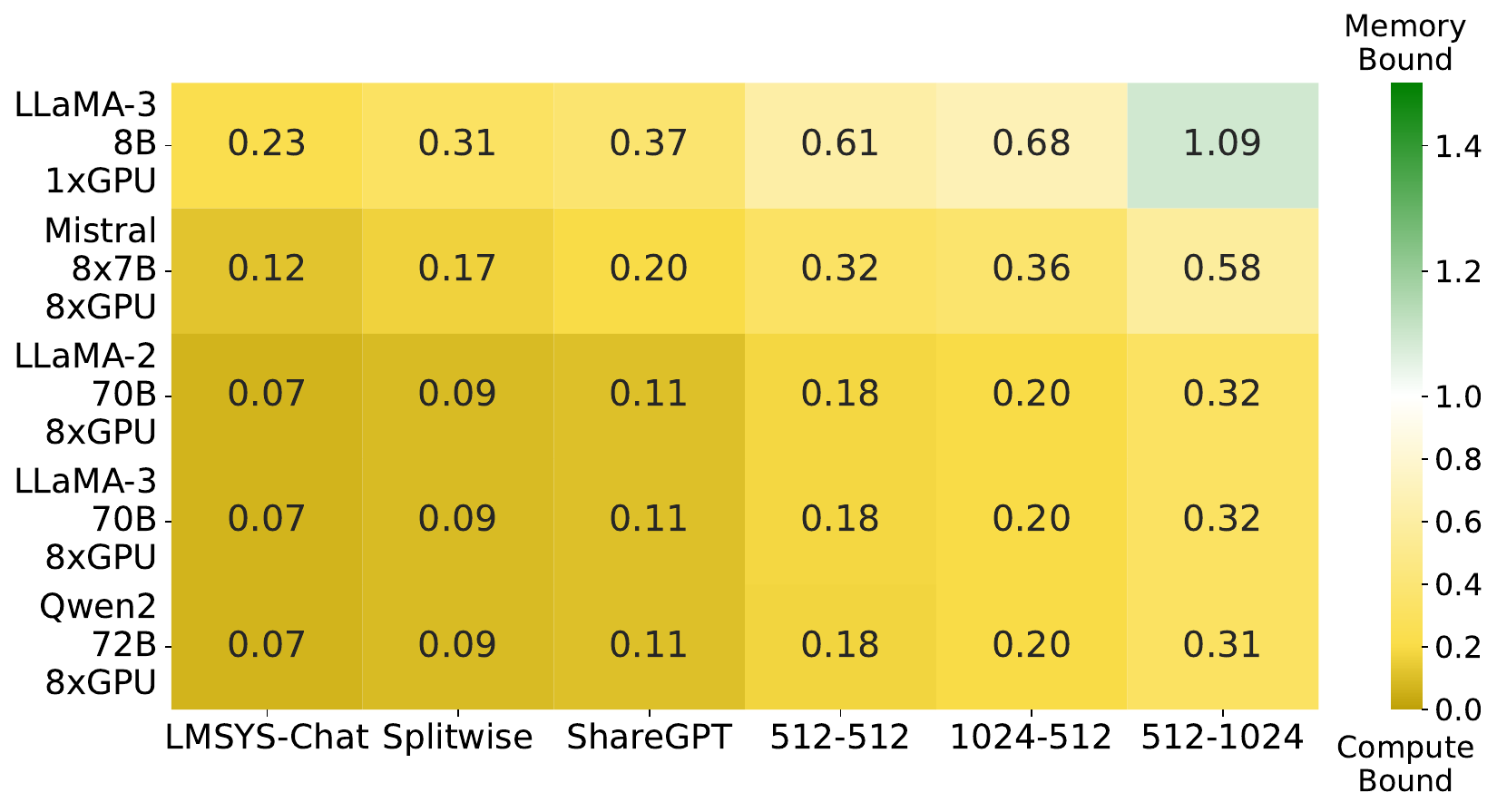}
    
    \caption{Comparison of compute time and memory time. The closer to yellow, the more compute-bound the workload is, whereas the closer to green the more memory-bound it becomes.}
    \label{fig:classification}
\end{figure}

For every GEMM in dense operations, $2B_{Dense}N_wK_w$ computations need to be performed on weights and activations, where $N_w$ and $K_w$ are dimensions of the weight matrices, and $B_{Dense}$ is the batch size for dense operations, which includes decode tokens from hundreds of requests and the prefill tokens from one or a few requests. Thus, the total compute for dense operations is $2B_{Dense}\sum{N_wK_w}$. Note that the $\sum{N_wK_w}$ term is the number of weight elements in all layers, which can be approximated using $P_{Model}$ (the parameters in the model), thus, we simplify the expression as $2B_{Dense}P_{Model}$. Therefore, latency from the compute perspective is: 
 \begin{align}
    T_{Compute} &\approx \frac{2B_{Dense} \cdot P_{Model}}{Compute} \label{eq:t-compute}
\end{align}

\noindent \textbf{Network.} Finally, for tensor parallelism, collective network communication primitives are needed to synchronize the results after performing operations in multiple GPUs over different shards of data. 
The size of the data transmitted with collective communication equals the size of inputs for dense operations, which have dimension [$B_{dense}, D_{model}]$. To support tensor parallelism, two AGs and one ARs (or 2 ARs) are required~\cite{megatron-lm}. An AR requires gathering outputs from other GPUs, performing local summation, and broadcasting results, thus it needs to transfer the activations twice, while an AG transfers activations once. Therefore, in both cases, we can estimate the total data movement for one GPU in bytes as $4\cdot B_{Dense}D_{model} S_{type}\cdot L$. Thus, the latency from the network perspective is:
\begin{align}
    \label{eq:net-latency}
    T_{net} \approx 4\cdot\frac{N_{GPU}B_{Dense}D_{model} S_{type} L }{NetBW}
\end{align}

\subsection{Classification of LLM Serving Workloads}\label{sec:classification}

We classify LLM serving workloads by comparing the latency of an iteration derived based on memory, compute, and network requirements. 
\begin{table*}[]
\centering
\caption{\blue{Characteristics of accelerator models across various vendors and release years.}}
\scalebox{0.8}{
\begin{tabular}{@{}cccccccccc@{}}
\toprule
Vendor &  Model  & \makecell{Release \\ Year} & \makecell{MemSize \\ (GB)} & \makecell{MemBW \\ (GB/s)} & \makecell{NetBW \\ (GB/s)} & \makecell{Compute \\ (FP16 GFLOP/s)} & $\dfrac{MemSize}{MemBW}$ & $\dfrac{Compute}{MemBW}$ & $\dfrac{NetBW}{MemBW}$ \\ \midrule
NVIDIA & V100       & 2017 & 16    & 900    & 300    & 125,000 & 0.018  & 139  & 0.33 \\ 
NVIDIA & A100       & 2020 & 40    & 1,555  & 600    & 312,000 & 0.026  & 200  & 0.39 \\
NVIDIA & A100       & 2021 & 80    & 2,000  & 600    & 312,000 & 0.040  & 156  & 0.30 \\
NVIDIA & H100       & 2023 & 80    & 3,352  & 900    & 989,000 & 0.024  & 295  & 0.268 \\
NVIDIA & H200       & 2024 & 96    & 4,800  & 900    & 989,000 & 0.020  & 206  & 0.19 \\
NVIDIA & B100       & 2024 & 120   & 8,000  & 1,800  & 1,800,000 & 0.015  & 225  & 0.23 \\
NVIDIA & B200       & 2024 & 120   & 8,000  & 1,800  & 2,250,000 & 0.015  & 281  & 0.23 \\ \hline
AMD    & MI250      & 2021 & 128   & 3,352  & 800    & 362,000 & 0.038  & 107  & 0.24 \\
AMD    & MI300      & 2023 & 192   & 5,300  & 1,024  & 1,307,000 & 0.036  & 246  & 0.19 \\
AMD    & MI325X     & 2024 & 256   & 6,000  & 1,024  & 1,307,000 & 0.043  & 218  & 0.17 \\ \hline
Intel  & Gaudi 2    & 2022 & 96    & 2,400  & 600    & 1,000,000 & 0.040  & 417  & 0.25 \\
Intel  & Gaudi 3    & 2024 & 128   & 3,700  & 1,200  & 1,800,000 & 0.035  & 486  & 0.32 \\ 
\hline
\makecell{NVIDIA} & \makecell{Ada 6000} & 2022 & 48 & 960 & 64 & 182,000 & 0.050 & 190 & 0.067 \\ 
\bottomrule
\end{tabular}
}
\label{tab:gpu-characteristics}
\end{table*}

First, we compare Equation~\ref{eq:net-latency} (network time) with Equation~\ref{eq:t-compute} (compute time). 
Their ratio can be calculated as 
$$
\frac{T_{Net}}{T_{Compute}} = \frac{2D_{model}L}{P_{model}} \frac{N_{GPU}Compute}{NetBW/S_{type}} 
$$

Note that $P_{model} \approx 12D_{model}^2 L$\footnote{$P_{model} = ((W_{K} + W_{Q} + W_{V}) + W_{O} + W_{U} + W_{G} + W_{D})L$ $\approx ((1/R_{GQA} + 1 + 1/R_{GQA}) D_{model}^2 + D_{model}^2 + D_{model}I_{model} + D_{model}I_{model} + I_{model}D_{model})L ) \approx (1+1+3.5+3.5+3.5)D_{model}^2L \approx 12D_{model}^2L$, where $I_{model}$ is intermediate dimension, which is typically 3.5x of $D_{model}$~\cite{llama3,mistral}.}, thus $\frac{2D_{model}L}{P_{model}} \approx 1/(6D_{model})$. For models that need multiple GPUs (where $T_{Net}$ is relevant), $D_{model}$ is typically larger than $4096$\cite{llama, llama3, qwen, mistral}, making the term $\frac{2D_{model}L}{P_{model}}$ smaller than $1\times 10^{-5}$. Due to the high bandwidth interconnect between data center GPUs (e.g., NVLink\cite{nvlink}, Infinity Fabric\cite{zhao2024forestcollefficientcollectivecommunications}), $\frac{N_{GPU}Compute}{NetBW/S_{type}}$ ranges from $1\times 10^{4}$ to $1\times 10^{5}$. Therefore,
this ratio is typically less than 1, indicating that the network is not the bottleneck. We illustrate this ratio in Figure \ref{fig:net-compute}. For large models and data center GPUs, the compute takes more time than network.

We then compare Equation \ref{eq:mem-latency} (memory time) and Equation \ref{eq:t-compute} (compute time) to finally determine the characteristics of the workload, namely, the following value 
\begin{align}
    \label{eq:t-ratio}
    T_R &= \frac{T_{Mem}}{T_{Compute}} \approx \frac{Compute}{MemBW} \frac{MemSize}{P_{model}} \frac{1}{2B_{dense}}
\end{align}

Modern models widely adopt Grouped Query Attention (GQA) as seen in LLaMA-3~\cite{llama}, NeMo~\cite{nemo}, and Qwen2~\cite{qwen}. In GQA, multiple attention heads share a \kv{}, which effectively means that given the same memory capacity, batches can contain more requests. Consequently, $B_{dense}$ tends to be larger. For example, when serving \lm{}-2 $70B$ on $8\times A100$ GPUs, the maximum batch size of decode requests is on the order of $1024$. Combined with the prefill tokens in the batch, $B_{dense}$ can reach $2048$ \blue{while a non-GQA model with the same size only can have $B_{dense} = 256$}. Thus, the ratio $T_R$ is less than 1, and compute becomes the main bottleneck. Additionally, modern models tend to have increasingly large model sizes $P_{model}$ (whereas the rest of the terms in $T_R$ are constants given a hardware specification), further causing $T_R$ to drop below 1.

Figure \ref{fig:classification} shows the values of $T_R$ with five representative models: \lm{}-3 8B \cite{llama3} with $1\times$A100, Mistral $8\times7$B \cite{mistral} with $8\times$A100, \lm{}-2 70B \cite{llama2} with $8\times$A100s, \lm{}-3 70B with $8\times$A100s, Qwen-2 72B \cite{qwen} with $8\times$A100s. As the heatmap figure shows, many of the workloads (Splitwise \cite{splitwise}, LMSYS-Chat-1M \cite{zheng2023lmsyschat1m}, ShareGPT \cite{sharegpt} as well as constant length inputs and outputs), are uniformly compute-bound, except in the case of the long decode (512-1024) workload on \lm{}-3 8B model, where the ratio $T_R$ is around $1$. 

\blue{
While the analysis is mainly conducted on an NVIDIA A100, the result remains similar for other accelerators. As shown in Table \ref{tab:gpu-characteristics}, compute/memory ratio and compute/network ratio are stable across vendors and generations, meaning the workload characteristics also remain largely unchanged.}

To summarize, our cost model suggests that modern LLM serving workloads operate in a compute-bound regime, which we validate empirically in the next section.

\subsection{Validation of the Cost Model}
\label{sec:validation}

We validated our cost model on \lm{}-2 70B serving with 8 NVIDIA A100 GPUs and a dense batch size of 2048 requests. We computed the GFLOPs, memory movement, and network traffic for each operation (Table~\ref{tab:validate}) and used these with our model to calculate $T_{compute}$, $T_{mem}$, and $T_{net}$ based on hardware specifications\footnote{One-way network bandwidth was used for $T_{net}$}. The longest estimated time $T_{op} = \max (T_{compute}, T_{mem}, T_{net})$ indicates the most constrained resource. To validate these results, we measure the running time of different operations. For most operations, $T_{op}$ aligns with profiling results. An exception is prefill attention, where the real measured time is larger due to the launching overhead of the associated kernels dominating the total time. The sums of all operations' $T_{compute}$, $T_{mem}$, and $T_{net}$ values show that compute is the most constrained resource, which aligns with our model's finding.

\subsection{Optimal Serving Throughput}
\label{sec:analysis-optimal}

Optimal throughput (measured in token/s/GPU) is achieved when the most limited resource, i.e. compute, is fully utilized. By applying Equation~\ref{eq:t-compute}, the system's optimal throughput is:

\begin{align}
    \label{eq:optimal-throughput}
    \mathrm{Throughput_{optimal}} &= \frac{B_{Dense}}{T_{Compute}} 
    = \frac{Compute}{2 P_{Model}}
\end{align}

In the compute-bound regime, the optimal throughput depends \textit{solely} on the aggregate computational capacity of the GPUs for the specific data type and the number of parameters in the model. Notably, other factors, including the GPU memory size, bandwidth, model data type, or the length of prefill and decode phases, do not have a significant impact on achieving optimal throughput. \blue{We measure the GPU compute capacity by profiling the state-of-the-art GEMM vendor library, CUTLASS\cite{Thakkar_CUTLASS_2023}}. For example, when serving \lm{}-2 70B on $8\times$A100 GPUs, \blue{the profiled peak $Compute$ is $280$ TFLOPS for FP16}, and the $P_{Model}$ is $70$B. Substituting these into Equation 5, yields to an optimal throughput of $1857$ tokens/s/GPU.
 
\begin{table}[]
\caption{Comparison of operation runtimes between cost model estimation and real-world measurements.} 
\label{tab:validate}
\scriptsize
\begin{tabular}{lrrrrrrr}
\hline
\multicolumn{1}{c}{Operation} &
  \multicolumn{1}{c}{\makecell{Compute\\(GFLOP)}} &
  \multicolumn{1}{c}{\makecell{Mem\\Load\\(GB)}} &
  \multicolumn{1}{c}{\makecell{Net\\Usage\\(GB)}} &
  \multicolumn{1}{c}{\makecell{Est.\\$T_{comp}$\\(ms)}} &
  \multicolumn{1}{c}{\makecell{Est.\\$T_{mem}$\\(ms)}} &
  \multicolumn{1}{c}{\makecell{Est.\\$T_{net}$\\(ms)}} &
  \multicolumn{1}{c}{\makecell{Real\\Time\\(ms)}} \\ \hline
KQV          & 27487.8              & 19.5                 & 0                    & \textbf{11.01} & 1.22           & 0              & \textbf{16.08}                \\
O            & 21990.2              & 16.1                 & 0                    & \textbf{8.81}  & 1.01           & 0              & \textbf{16.01}                   \\
UG           & 153931.6             & 96.6                 & 0                    & \textbf{61.67} & 6.04           & 0              & \textbf{69.92}                \\
D            & 76965.8              & 49.7                 & 0                    & \textbf{30.84} & 3.11           & 0              & \textbf{34.96}                \\
DecAttn  & 3665.9               & 462.2                & 0                    & 1.47           & \textbf{28.89} & 0              & \textbf{35.60}                 \\
PfAttn & 916.3                & 2.1                  & 0                    & \textbf{0.37}  & 0.13           & 0              & \textbf{4.56}                 \\
Net     & 18.8                 & 75.2                 & 75.2                 & 0.01           & 4.70            & \textbf{31.33} & \textbf{47.92}                \\ \hline
Total    & \multicolumn{1}{l}{} & \multicolumn{1}{l}{} & \multicolumn{1}{l}{} & \textbf{114.17}         & 45.09          & 31.33          & \multicolumn{1}{l}{} \\
\end{tabular}
\end{table}

\subsection{Gap to Optimal Throughput}
\label{sec:analysis-existing}

As shown in \fig{fig:offline-throughput}, existing serving systems fall far short of achieving optimal throughput. 
When serving offline requests, vLLM, DeepSpeed-FastGen, and TensorRT-LLM achieve only 22.0\%, 22.9\%, and 37.8\% of the optimal throughput, respectively.

\begin{figure*}
    \centering
    \includegraphics[width=\textwidth]{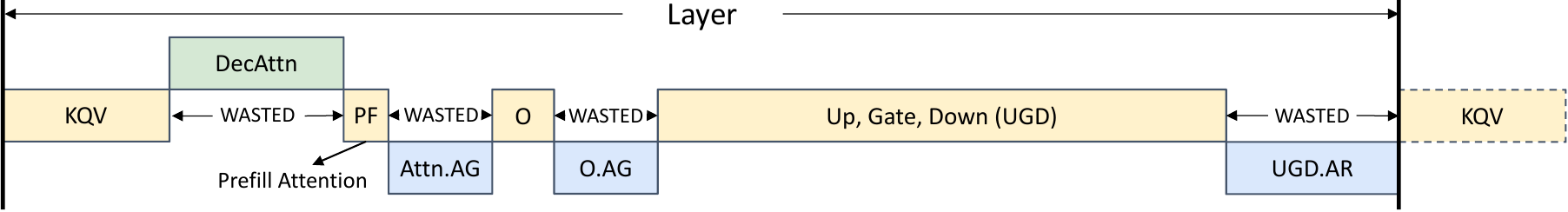}
    \caption{Execution pipeline of existing systems. The green, yellow, and blue operations correspond to memory-, compute-, and network-bound operations.  Operations in the previous and next layer are denoted by dotted borders. "WASTED" shows the stages in the pipeline where the most constrained resource, compute, is underutilized. Small operations (i.e. layernorm, activation, etc.) are omitted for simplicity.}

    \label{fig:exist-pipe}
\end{figure*}

The main reason why existing serving systems fall significantly short of achieving optimal throughput is poor utilization of GPU resources. \fig{fig:exist-pipe} shows the execution flow of transformer operations within a GPU, using existing serving frameworks like SGLang \cite{zheng2024sglangefficientexecutionstructured}, vLLM \cite{vllm}, DeepSpeed-FastGen \cite{deepspeed-fastgen}. These frameworks execute compute bound operations, decode attention, and network operations sequentially on a single large batch in a device. This sequential execution flow forms multiple pipeline bubbles (denoted as "WASTED" in \fig{fig:exist-pipe}), which cause the most constrained resource, compute, to be underutilized.

\subsection{\NewParall{}}
\label{sec:analysis-intra}

Ostensibly, the advantage of existing serving frameworks when using a single large batch in a GPU is that they need to load model weights fewer times than if there were multiple smaller batches and corresponding operations.

However, the cost model and its validation that we outlined earlier in this section provides a valuable insight: Since modern models are largely compute-bound (by a factor of more than 2 as shown in \fig{fig:classification} and Table~\ref{tab:validate}), creating and processing smaller batches may be justified, especially if compute utilization can be increased.

These observations motivate intra-device parallelism through \nano{}ing, which underlies the design of \pllm{}. \blue{\pllm{} creates multiple \nanoop{}s, which are the same operations as the original operations, but operate on finer-grained batches that we call \nano{}es. For example, for the original Up projection operating on a batch size of 2048, \pllm{} could create 2 nano-operations UP1 and UP2, operating on batches in ranges 0-768 and 768-2048, respectively, while performing the exact same projection.} Different \nano{}es do not have data dependencies. Therefore, \nanoop{}s that use different resources---compute, memory, network---can operate on different \nano{}es in parallel within a device to maximize the utilization of the most constrained resource,  compute. Despite requiring repeated loading of model weights, we demonstrate that \pllm{} provides significantly more throughput than existing frameworks (1.91$\times$ on average), further closing the gap to optimal serving throughput (\S\ref{sec:evaluation}).

\section{Design}

We introduce \pllm{}'s key components: the \auto{} engine for automatic \nano{} pipeline computation (\S\ref{sec:design}) and the LLM serving runtime for pipeline execution (\S\ref{sec:implementation}).

\subsection{Automated Pipeline Search}
\label{sec:design}

Deciding each \nanoop{}'s resource allocation, launching kernels, and scheduling with \newparall{} is challenging due to model and hardware diversity. To address this, \pllm{} proposes \auto{}, which uses mixed integer linear programming to obtain the \nano{} overlapping strategy, with a two-stage approximation for reducing the search space.
We first outline \auto{} prerequisites including kernel profiling and interference modeling (\S\ref{subsec:kernel-profile}), then detail its two stages: pipeline structure search (\S\ref{subsec: pipeline-search}) and GPU resource allocation (\S\ref{subsec: resource-search}). Finally, we provide example pipelines for popular models (\S\ref{subsec: case-study}).

\subsubsection{Kernel Profiling and Interference Modeling}
\label{subsec:kernel-profile}
Understanding the performance characteristics of different operations is key to automated pipeline search. For this, \pllm{} profiles GEMM kernels (used for compute-bound operations such as KQV generation), GEMV kernels (used in memory-bound operations as in decode-attention), and network kernels (used in network-bound operations such as AR and AG). 

\textbf{Determining the maximum dense batch size}
Given a specific model architecture and user input statistics (as described in \S\ref{sec:costfactors}), \pllm{} calculates the maximum \textit{dense batch size} that can fit into the GPU memory (e.g., $2048$ for $8\times A100$ serving \lm{}-2 70B). This dense batch size serves as an input to \auto{}, which sets an upper bound of shapes in profiling.

\textbf{Profiling interference-free kernels}
\pllm{} then determines the interference-free execution time of kernels, where each operation exclusively uses the GPU. \pllm{} profiles kernels with discrete input batch sizes from $128$ to the dense batch size in multiples of $128$, as $128$ is a hardware-friendly shape for GEMM tiling. 
\pllm{} explores all possible kernel implementations varying the number of thread blocks, the number of warps, and tile size for GEMM, GEMV, and network kernels to identify the implementation that provides the shortest execution time. The profiling outputs a mapping from (kernel, batch size) to its best implementation.

\textbf{Profiling kernels with interference}
Ideally, we would like to independently assign each \nanoop{} a fractional allocation of a single GPU's compute, memory, and network bandwidth.
We call this fraction $R_{physical}$.
However, when two or more kernels execute in parallel on a device, they slow down compared to running separately. We refer to this as \textit{kernel interference}, which is attributed to kernels competing for GPU hardware resources such as execution units, caches, and memory bandwidth~\cite{orion}.
Unfortunately, kernel interference is unpredictable, as NVIDIA GPUs do not give explicit control over compute, memory, and network bandwidth.
Thus, we cannot directly control $R_{physical}$.

Instead, we use GEMM performance as a proxy $R$ for $R_{physical}$ and model pairwise kernel interference.
We define $R$ in a GEMM-centric way since the compute-bound operations dominate and optimizing compute is the goal of \pllm{}.

In particular, we profile pairwise kernel interference patterns by measuring actual performance when overlapping between a compute kernel A and another kernel B.
For example, if when overlapped, kernel A achieves 40\% performance compared to its best performance when run individually, then we say that $R_A$=0.4.
We then assume that the overlapped kernel is given the remaining resources, so we set $R_B=1-R_A=$0.6.

Then, we model the interference by measuring the performance loss in the non-compute kernel B when $R_A$ resources are allocated to kernel A.
When overlapping A and B, we normalize A and B's performance to their peak performance when run individually.
We call this value $P$.
For the compute kernel A, $P_A$ is equivalent to $R_A$ by definition, while $P_B$ captures the memory or network utilization when $R_A$ resources are assigned to the compute kernel.

Intuitively, this process establishes an exchange rate between $R$ compute utilization to $P$ memory or network utilization.
The overall kernel interference model can then be captured in a table such as Table \ref{tab: occu}, which maps $R$ to $P$ for each kernel type.

Nonetheless, profiling the mapping between \(R\) and \(P\) remains challenging due to the vast number of possible kernel combinations.
While each operation may have only a few hundred kernel implementations, the combinations between different implementations of overlapping kernels exponentially expand the profiling space, resulting in millions of possible configurations. This immense complexity makes exhaustive exploration infeasible.

\pllm{} reduces the profiling space by limiting the thread block numbers for GEMV and network kernels to values between \(8\) and \(128\) (in steps of \(8\)), as \(128\) blocks are sufficient to saturate their performance.
It also excludes less efficient GEMM kernels which have longer execution times while using more thread blocks. 
\pllm{} focuses on pairwise interference analysis---compute-memory and compute-network---rather than examining three kernel types simultaneously and assumes that the \(R\) to \(P\) mapping profiled with pairwise interference holds when overlapping three kernels.
\begin{figure}[t]
    \centering
    \includegraphics[width=\columnwidth]{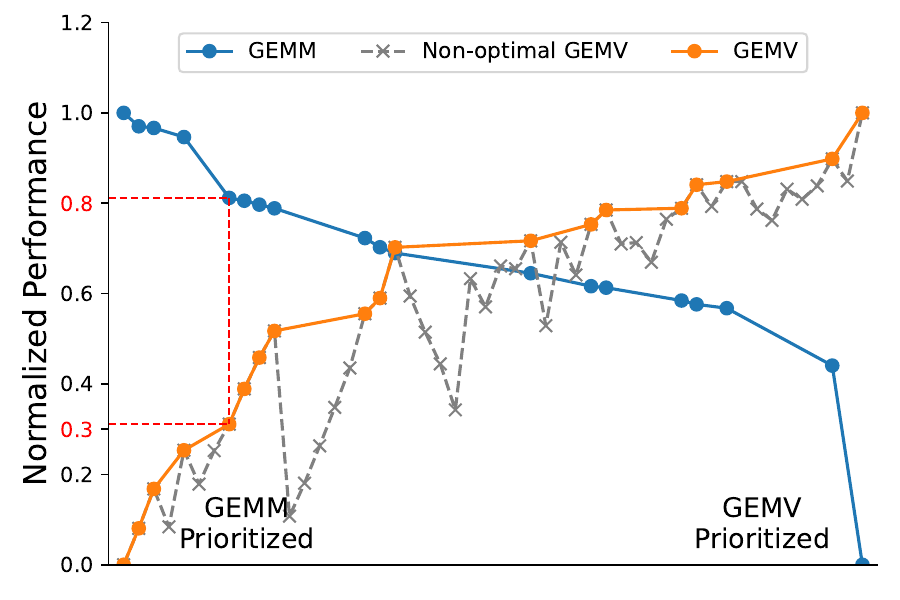}
    \caption{Interference characteristics between GEMM and GEMV kernels. The points on the x-axis correspond unique GEMM-GEMV implementation pairs. The y-axis denotes the GEMM and GEMV kernels' normalized performance \(P\).}
    \label{fig:interference}
\end{figure}

\begin{table}[]

\caption{Performance \(P\) of GEMV and network kernels with different resource utilization $R$.}
\begin{tabular}{cccccccc}
\multicolumn{1}{c}{\multirow{2}{*}{Operations}} & \multicolumn{7}{c}{Resource Utilization ($R$)} \\ \cline{2-8} 
\multicolumn{1}{c}{}                            & 0  & 0.1  & \red{0.2}  & …  & 0.8  & 0.9  & 1  \\ \hline
\makecell{GEMM\\(by definition)}                                         & 0  & 0.1  & 0.2  & …  & 0.8    & 0.9    & 1  \\
GEMV                                          & 0  & 0.2  & \red{0.3}  & …  & 0.85  & 0.95  & 1  \\
Network                                         & 0  & 0.3  & 0.5  & …  & 0.9    & 1    & 1 
\end{tabular}
\label{tab: occu}
\end{table}

Figure \ref{fig:interference} shows \pllm{}'s profiling results of GEMM-GEMV kernel pairs ($\sim 100$ pairs, after the aforementioned simplifications) on an A100 GPU with GEMM shape \((M,N,K) = (384, 4096, 4096)\), GEMV batch size \(384\), and sequence length \(1024\). These pairs are then sorted by descending GEMM performance, as shown in Figure \ref{fig:interference}. For instance, a normalized performance of $0.8$ means the kernel’s execution time is $1/0.8$ times longer than the fastest implementation. 
Pairs with higher interference on GEMM but worse GEMV performance (grayed-out in Figure \ref{fig:interference}) are discarded. This process identifies the best-performing kernel combinations at various GEMM-GEMV performance trade-off points.

Using interference profiles, \auto{} generates an \textit{resource mapping table} (Table \ref{tab: occu}). This table quantifies GEMV and network kernel performance (\(P\)) as functions of resource utilization (\(R\)).
For example, as depicted by the dotted red line in Figure \ref{fig:interference}, achieving $0.3$ GEMV performance requires sacrificing $0.2$ GEMM performance (from $1$ to $0.8$). 
This implies a non-linear trade-off: $0.2$ GEMM performance is exchanged for $0.3$ GEMV performance, indicating that \(R_{\text{GEMV}} = 0.2\) corresponds to \(P_{\text{GEMV}} = 0.3\).

A sensitivity analysis of all GEMM shapes in the evaluated LLM models and 64 batch size combinations reveals that the \(R\) to \(P\) mapping is consistent, with a standard deviation within \(5\%\) of the mean.
Consequently, Table \ref{tab: occu} can be used to map \(R\) to \(P\) for all GEMV and network kernel implementations in subsequent \auto{} analysis.

\subsubsection{\Auto{} Stage I: Pipeline Structure Search}
\label{subsec: pipeline-search}
In the first stage, \auto{} takes as input the dense batch size, the operation dependencies (defined in the PyTorch implementation of a given model architecture), and the profiling of interference-free kernels, and then produces as output, the number, batch size, and order of each \nanoop{} using mixed integer linear programming (MILP). Stage I does not factor in the interference between \nanoop{}s to simplify the problem, which is covered by Stage II. The MILP problem is subject to the constraints described below.

\textbf{Optimization objective.} The main objective of MILP is to remove pipeline bubbles (denoted by wasted in \fig{fig:exist-pipe}) for compute operations to minimize execution time.

\textbf{Constraints on the number of \nanoop{}s.} 
To overlap resources, each operation needs to be split into at least two \nanoop{}s, each operating on an independent \nano{}. As splitting into \nanoop{}s reduces the batching effect, the number of \nanoop{}s should be minimized. Thus, \auto{} begins its search by dividing all operations into two \nanoop{}s and formulating the MILP problem to identify the batch size, execution time, and ordering of the \nanoop{}s. If the temporary best solution has bubbles for compute, \auto{} increases the number of \nanoop{}s for operations near the bubble to improve resource utilization until MILP cannot produce better solutions.

\textbf{Constraints on batch sizes and execution times.} \Auto{} forces the batch sizes to be chosen from $128$ to dense batch size and uses the interference-free execution time of a \nanoop{} for a given batch size, as computed in \S\ref{subsec:kernel-profile}.

\textbf{Constraints on dependencies.}
The dependencies of \nanoop{}s are determined by their parent operations (present in baseline implementations, e.g., in pytorch) and their input batches.  Two \nanoop{}s are dependent if and only if their parent operations are dependent (e.g., an O-projection depends on an attention operation) and their input batches intersect (e.g., ranges 0–255 and 128–383). 

\textbf{Constraints on overlapping.}
Overlapping kernels constrained by the same resource (e.g., compute) is unhelpful for utilization, so \auto{} restricts the pipeline to overlap only operations constrained by different resources.

\textbf{Constraints on operation transformations.}
Some network collectives have equivalent transformations with different performance characteristics and dependencies. For example, an AG can be converted into an AR with different weight partitioning approaches~\cite{megatron-lm,alpa}. \pllm{} explores all of these alternative transformations for network \nanoop{}s and searches for the shortest pipeline design.

\blue{Due to the large search space, searching for a optimal pipeline would take hours or even days. However, a practical pipeline can be found in approximately 10 minutes by prioritizing searching a feasible solution instead of finding a provably-optimal solution.}

\subsubsection{\Auto{} Stage II: Refining the Pipeline}
\label{subsec: resource-search}

The pipeline derived in the first stage does not factor in kernel interference, which is impractical for real deployment. In the second stage, \auto{} refines this pipeline by considering kernel slowdown due to interference.
This stage formulates a separate MILP problem, where the numbers, batch sizes, and ordering constraints of \nanoop{}s remain the same (i.e., outputs of the first stage). 
\Auto{} explores various resource utilization (\(R\)) allocations to individual \nanoop{}s, and maps \(R\) to \(P\) using Table \ref{tab: occu} derived in interference profiling.
A specific resource utilization value \(R\) directly corresponds to kernel implementations as determined in \S\ref{subsec:kernel-profile}. The goal of the second stage is again to minimize pipeline execution time.

\textbf{Constraints on GPU resource utilization.} As discussed in \S \ref{subsec:kernel-profile}, concurrent kernels should compete for in total of $1.0$ of GPU resource utilization \(R\). Thus, \auto{} forces the sum of $R$ at any given time to be less than or equal to $1.0$.

\textbf{Constrains on execution times.} Given the resource utilization $R$ for a \nanoop{}, \auto{} computes the execution time of the operation using $D_{\text{best}} / P$, where $D_{\text{best}}$ represents the best execution time without interference and $P$ is derived using Table \ref{tab: occu}.

\blue{The auto-search is performed only when the model architecture or workload (input length, output length) undergoes significant changes. Compared to the long-running deployment times, the search time is negligible.}

\subsubsection{Example Pipelines}
\label{subsec: case-study}

\begin{figure*}
    \includegraphics[width=0.9\textwidth,center]{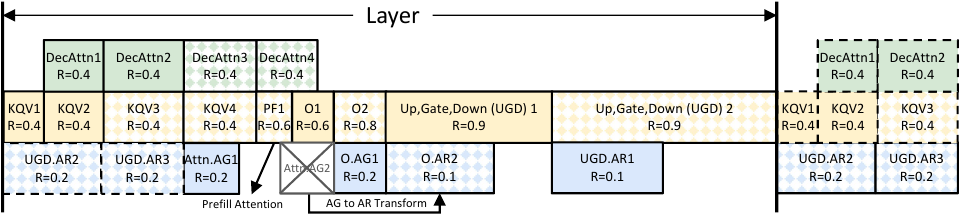}
    \caption{Execution pipeline of \lm{}-2 70B, automatically generated by \pllm{}. The solid background and shaded background represents input batch 0-768 and 768-2048, respectively. $R$ stands for resource utilization.
    By overlapping the compute-, memory-, and network-intensive operations, \pllm{} increases compute utilization and improves the serving throughput.
    }
    \label{fig:compute-pipe}
\end{figure*}

\textbf{70B pipeline} We apply \auto{} to 70B-scale models, including \lm{}-2 70B, \lm{}-3 70B, Qwen2.5-72B, and Deepseek-67B. Compared to \lm{}-2 70B, \lm{}-3 70B features a larger vocabulary size of $128$K, \blue{which increases the sampling time}; Deepseek-67B has a slightly different number of layers and hidden dimension ($D_{model}$), Qwen2.5-72B introduces bias in KQV generation. However, none of these adjustments significantly change the performance characteristics, leading to similar schedules. We provide the \pllm{}-generated pipeline for \lm{}-2 70B as an example in \fig{fig:compute-pipe}.

Since three resources (compute, memory, and network) overlap at the beginning of a decoding layer (i.e., KQV generations), \auto{} ends up using $4$ \nanoop{}s for this part, to reduce pipeline bubbles. Here, the decode attention operation has resource utilization $0.4$ (i.e. reduces $40\%$ GEMM performance) and reaches 80\% of maximum decode attention performance. For the remainder of the pipeline, GEMM operations are prioritized, and only two \nanoop{}s are used.

\textbf{8B pipeline} 8B models (e.g. Llama3-8B) do not require network operations as they fit in a single GPU. Thus \auto{} splits all operations into two \nanoop{}s, with decode attention overlapping with the Up Gate Down projection.

\textbf{MoE pipeline} 
Compared to 70B models, MoE models have different hidden dimensions and number of layers. Due to the imbalance between experts\cite{mistral}, \pllm{} uses tensor parallelism for MoEs, in which FFN is implemented using grouped-GEMM. FFN also involves an additional gate routing operation. 
\Auto{} works similarly for an MoE model and automatically produces an efficient pipeline~(\S\ref{sec:evaluation}).

\subsection{\pllm{} Runtime}
\label{sec:implementation}

We now explain how \pllm{} executes the auto-generated pipeline by asynchronous scheduling (\S\ref{sec:runtime-async}), and \kv{} SSD offloading for multi-round conversations (\S\ref{sec:runtime-offload}).

\subsubsection{Request Scheduling}
\label{sec:runtime-async}

\textbf{Batch formation.} 
\blue{
\pllm{} assumes that auto-scaling, workload balancing, and priority-aware routing are managed by the control plane externally~\cite{theaibrixteam2025aibrixscalablecosteffectivelarge,patke2025hierarchicalautoscalinglargelanguage,jain2025intelligentrouterllmworkloads}. A \pllm{} instance assumes an abundance of requests and treats each request with equal priority. When requests are not abundant, the control plane should reduce the number of \pllm{} instances to maintain a sufficiently large per-instance batch size. Note that while the $B_{dense}$, i.e. the token batch size for GEMMs, is on the order of thousands  (e.g., 1 prefill request of 2048 tokens and 256 decode tokens), this represents the combination of prefill and decode tokens. Therefore, the request batch size (e.g., 1 prefill and 256 decode requests) is on the order of hundreds, which is easy to attain in real-world large-scale serving.
}

\blue{When forming a batch, \pllm{} prioritizes unfinished decode requests and chunks prefill requests at a token granularity following SarathiServe~\cite{sarathi} to exactly fill the remaining capacity of a pre-selected best-performing dense batch. This effectively reduces the tail latency as dense operations are performed over consistent batch sizes across iterations.} \blue{Initially, the global batch contains only prefill requests (requests in prefill stages). As these requests complete the prefill stage, they transition to decode requests within the batch, and new prefill requests are introduced to maintain the fixed token batch size.  As decode requests increase, the prefill token budget decreases, causing prefill requests to slow down. When decode requests begin completing, this frees space for more prefill tokens again, speeding up prefill. Thus, the number of decode requests and prefill requests will be stabilized.
}

To optimize GPU memory usage and avoid running out of memory, \pllm{} predicts peak future memory usage based on the status of user requests. \pllm{} tracks decoded tokens per request, estimates completion time using average decode length, and prefills new requests only if predicted memory usage stays within GPU limits. If the GPU runs out of memory, \pllm{} offloads a request to the CPU and reloads it once memory is available without recomputation.

\textbf{Asynchronous scheduling.} Batch formation, including estimating memory usage, scheduling new requests, retiring finished requests, and adjusting the page table for PagedAttention~\cite{vllm}, consumes a non-negligible amount of time on the CPU side~\cite{cpuoverhead}. In most serving frameworks~\cite{zheng2024sglangefficientexecutionstructured,vllm}, \textit{only after} the GPU executes one iteration, the scheduler on the CPU is able to detect $EOS$ tokens, remove the finished request from the batch, and refill the batch with new requests. However, GPU is under-utilized during this time. To avoid this waste, \pllm{} asynchronously schedules batch formation in parallel to the GPU executions. At any iteration $i$, \pllm{} forms the batch for the next iteration \textit{before} the end of the current iteration. This means that \pllm{} cannot detect the $EOS$ tokens generated at iteration $i$. After launching iteration $i+1$, \pllm{} forms the batch for cycle $i+2$, detects the $EOS$ token from iteration $i$, and removes finished requests. Fortunately, since the average decode length surpasses $100$ for typical workloads (See Table~\ref{tab:datasets}), the overhead of one extra decode token is negligible ($<1\%$), given the benefit of hiding the batch formation overhead.

\subsubsection{\kv{} Management}
\label{sec:runtime-offload}
To efficiently serve multi-round applications, \pllm{} offloads the \kv{} of running requests to a hierarchical cache consisting of host CPU memory and SSDs, and loads a prior round's \kv{} when a new round arrives.

\textbf{Simultaneous offloading.}
Instead of waiting for requests' completion, \pllm{} offloads the \kv{} of tokens directly after the KQV generation in each transformer layer before appending them to the \kv{}. The KV vectors output by KQV generation are naturally contiguous and the data size of the offload is balanced across iterations by design as \pllm{}'s dense batches remain stable. As a result, the host and the GPU hold the same copy of on-the-fly requests' \kv{}. To minimize the offloading overhead, \pllm{} performs the device-host copy---a GPU-initiated copy operation that uses minor GPU resources---during compute-bound operations in FFN. Moreover, \pllm{} reduces offloading time via NUMA-aware thread-binding.

\textbf{Host \kv{} management.}
\pllm{} uses the LRU policy to manage the hierarchical cache of CPU memory and SSDs. For e.g., \kv{} are evicted to SSDs when the CPU reaches its memory limit and retrieved from either CPU or SSD to initiate the loading process when the next round conversation of a request comes in.

\textbf{\kv{} loading and scattering.} 
\pllm{} uses PagedAttention~\cite{vllm}, therefore, the pages of the \kv{} reside in the GPU memory in a fragmented manner. In order to avoid copying to fragmented GPU page destinations, \pllm{} first copies the data to a contiguous space on GPU and then scatters the pages to their destinations. This achieves $7\text{-}10\times$ higher bandwidth for host-to-device copy.

\section{Implementation}
We implement \pllm{} for NVIDIA GPUs. \pllm{} consists $\sim$10K lines of CUDA and $\sim$6K lines of Python.

Given GPU resource requirements determined by \auto{}, \pllm{} knows which kernels to launch for each \nano{} to achieve a given utilization level, based on its profiling results. \pllm{} launches \nanoop{}s respecting the ordering constraints from \auto{} and enforces ordering dependencies using CUDA events.

\section{Evaluation}
\label{sec:evaluation}

\begin{figure}[t]
  \centering
  \subfloat{
    \includegraphics*[width=\columnwidth]{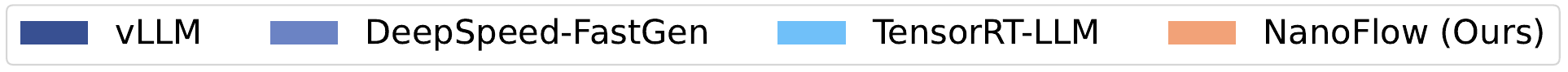}
  }
  \setcounter{subfigure}{0}
  \subfloat[LLaMA-2-70B, 8 GPU, TP=8, Constant Input \& Output Length]{\includegraphics[width=\columnwidth]{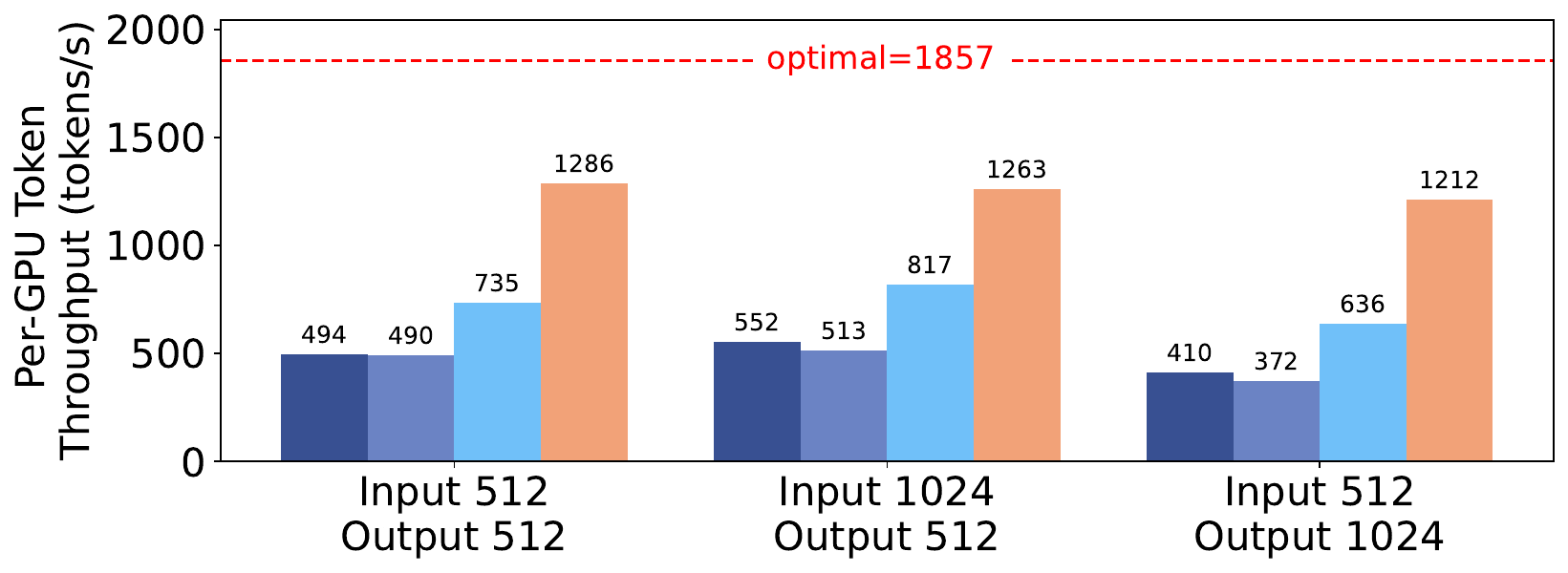}}\label{fig:eval-offline-70b}
  \subfloat[LLaMA-2-70B, 8 GPU, TP=8, Input \& Output Length from Dataset]{\includegraphics[width=\columnwidth]{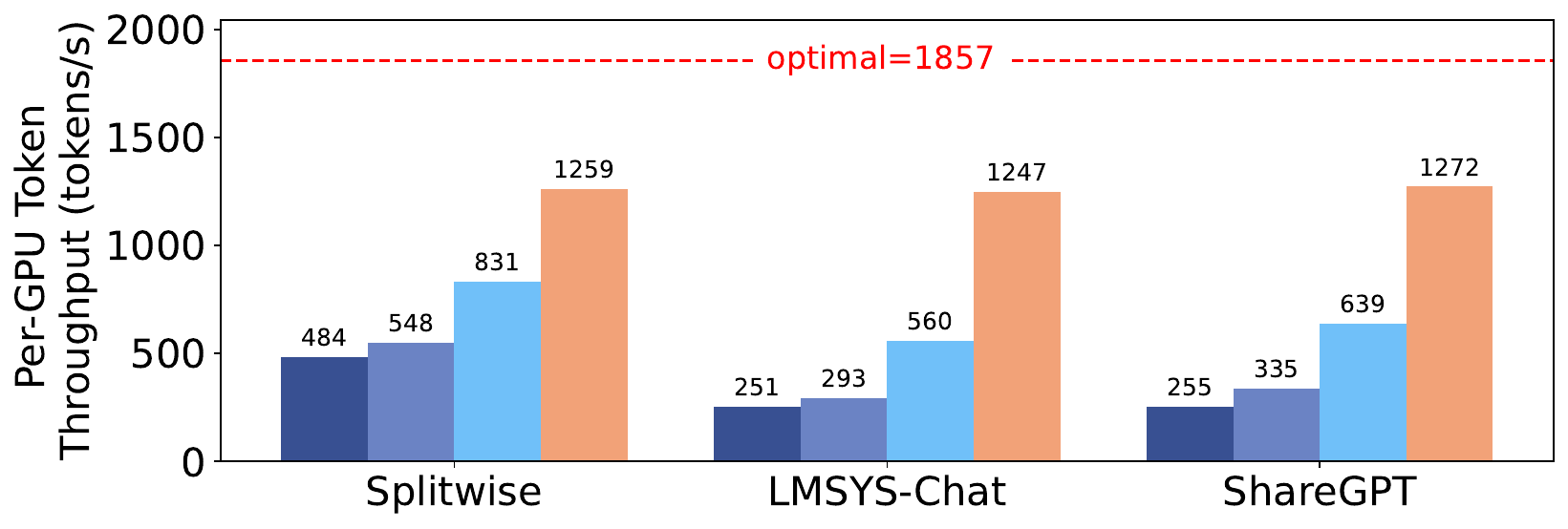}}\label{fig:eval-offline-dataset-70b}
  \caption{Offline throughput comparison. \pllm{} outperforms all baselines for all the workload settings. TP stands for the number of GPUs used with tensor parallelism.}\label{fig:offline-throughput}

\end{figure}

We now mainly answer the following questions:
\begin{itemize}[noitemsep, topsep=0pt]
    \item How much \pllm{} improves serving throughput compared to existing systems, and how does its throughput compare to the optimal serving throughput? (\S\ref{sec:eval-offline})
    \item What is the average serving latency under \pllm{} for different request rates, and how is it distributed? (\S\ref{sec:eval-latency})
    \item How do the various techniques proposed in \pllm{} contribute to the end-to-end throughput? (\S\ref{sec:eval-ablation})
    \item What is the compute, memory and network resource usage pattern of \pllm{}?
    (\S\ref{sec:eval-resource})
    \item How does \pllm{} improve performance when applied to different popular LLMs? (\S\ref{sec:eval-porting})
\end{itemize}

\subsection{Experiment Setup}

\textbf{Hardware.} We run our experiments on $8\times$ A100 80GB SXM GPUs interconnected via NVLink. 

\textbf{Models.} We provide a detailed evaluation of \pllm{} using the \lm{}-2-70B model, one of the most widely-used open-source LLMs. We also demonstrate the applicability of \pllm{} by evaluating it on 5 other representative LLMs. \blue{We use FP16 weights and activations for all models, as this is standard for data center-scale inference.}

\textbf{Baselines.} 
We consider three widely-used serving frameworks as baselines. 
\begin{figure*}[!t]
  \centering
  \subfloat{
    \includegraphics*[width=0.5\textwidth]{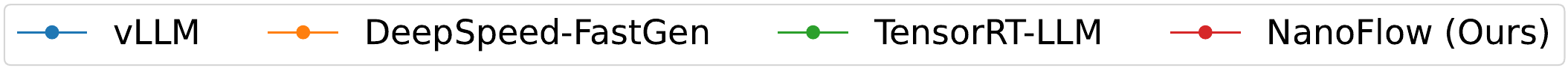}
  }

  \setcounter{subfigure}{0}
  \subfloat[Splitwise]{
  \includegraphics[width=0.325\textwidth]{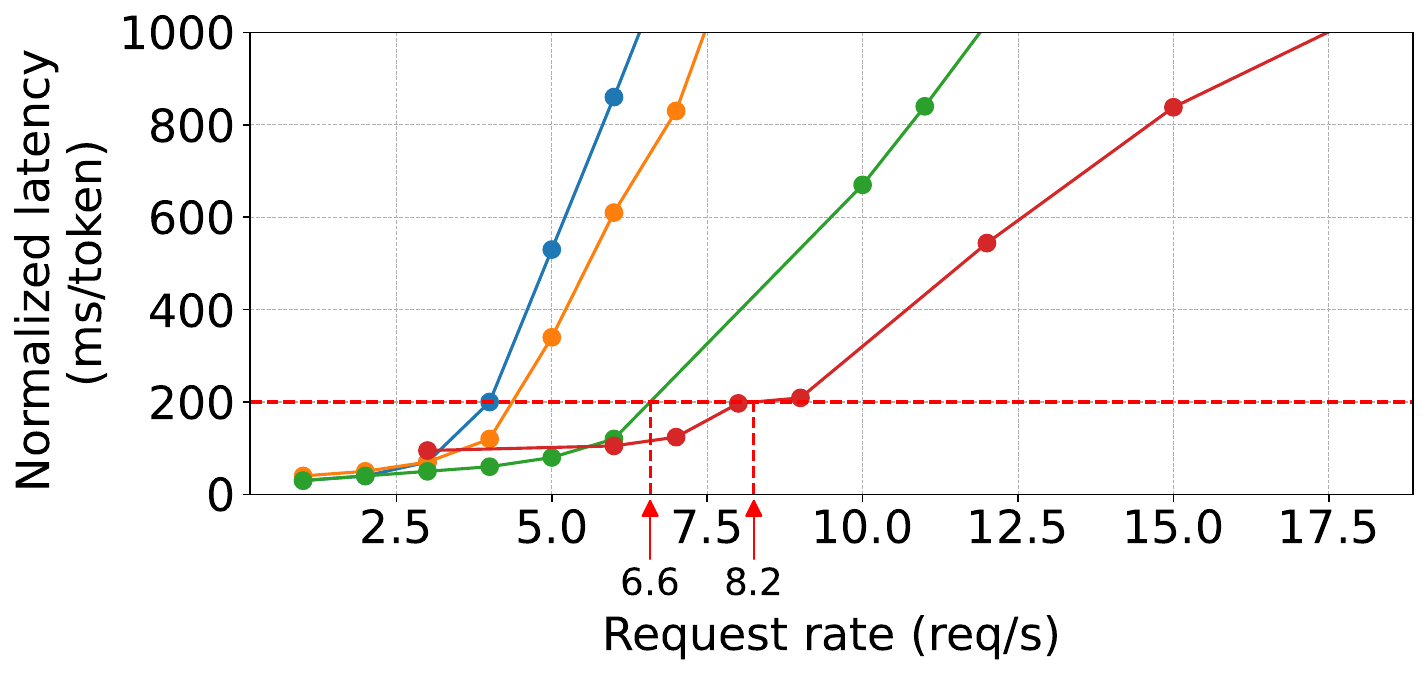}}\label{fig:eval-online-70b-splitwise}
  \subfloat[LMSYS-Chat-1M]{
  \includegraphics[width=0.325\textwidth]{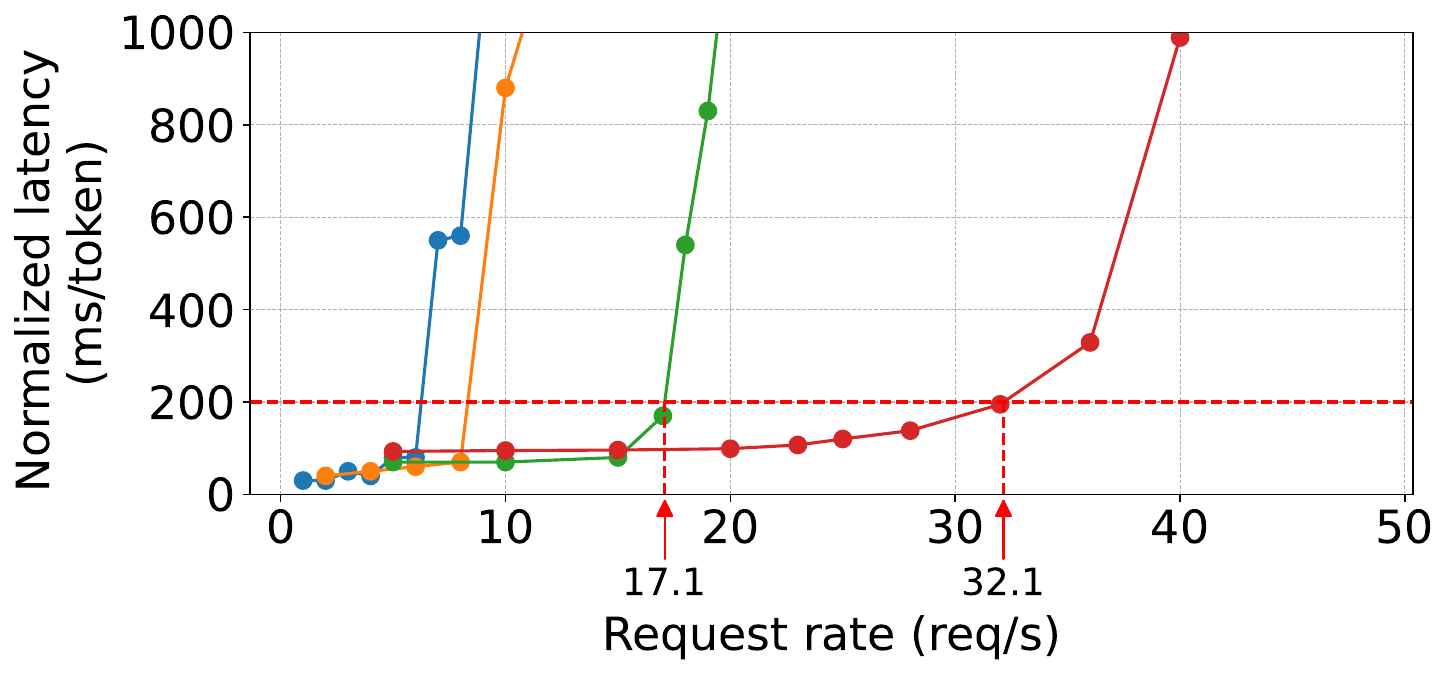}}
  \subfloat[ShareGPT]{
  \includegraphics[width=0.325\textwidth]{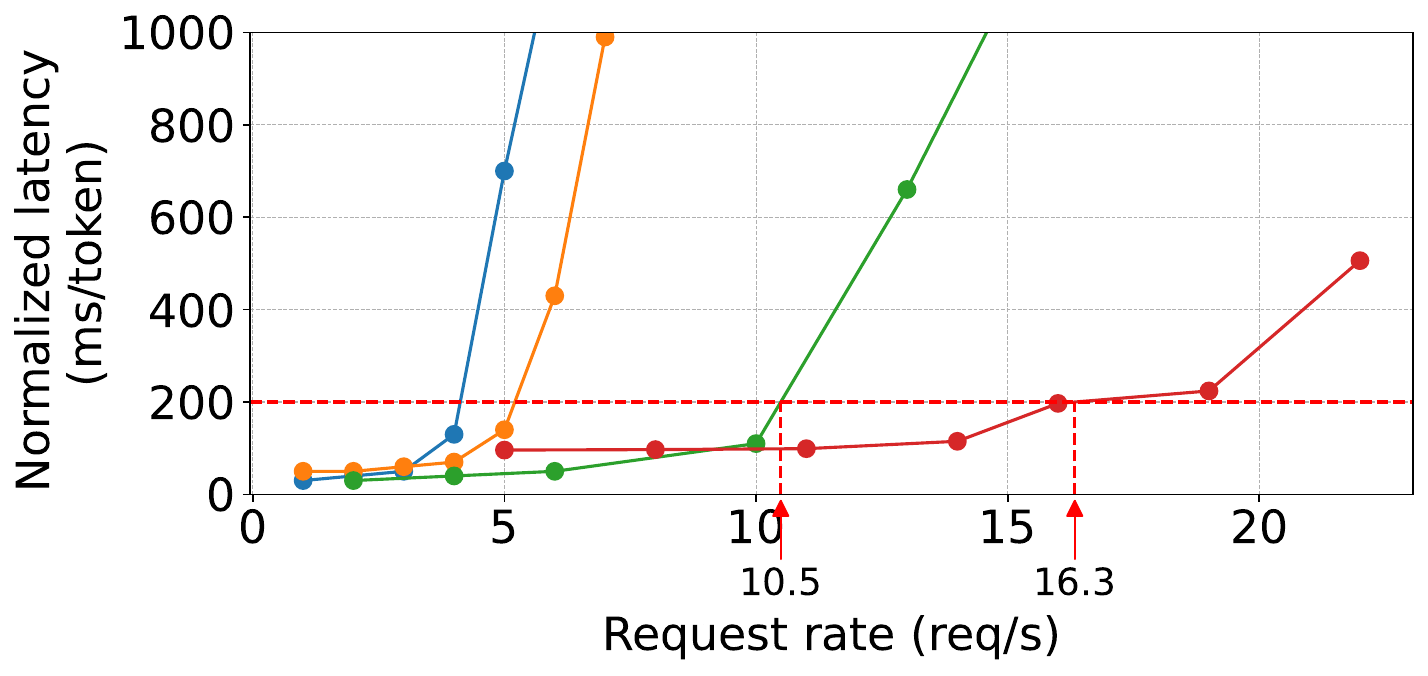}}
  \caption{Latency comparison. The x-axis shows the number of incoming requests per second and the y-axis shows the normalized latency. \pllm{} handles higher request within 200ms SLO constraints. }\label{fig:online-latency}
\end{figure*}

vLLM\footnote{v0.5.3.post1 (commit ID 38c4b7e)}~\cite{vllmgithub, vllm} is a state-of-the-art serving system delivering high throughput. \blue{vLLM implements pagedAttention for increasing GPU memory utilization, as well as the chunked prefill for higher GPU utilization.}

DeepSpeed-FastGen\footnote{v0.2.3 (commit ID 429bc5c)}~\cite{deepspeedgithub, deepspeed-fastgen} is a serving framework developed by Microsoft. 
It dynamically composes prefill with decode requests to ensure that the engine is operating in a high throughput regime. We vary the \texttt{max-ragged-batch-size} to tune the batch size for highest throughput.

TensorRT-LLM\footnote{v0.8.0 (commit ID 5955b8a)}~\cite{tensorrtgithub, tensorrtllm} is a high-performance LLM inference engine built upon NVIDIA's TensorRT SDK. We set \texttt{max-num-tokens} by calculating the maximum capacity for the \kv{} in the GPU memory. We also enable paged \kv{} and dynamic batching optimizations when compiling.

\begin{table}[]
\begin{tabular}{c|c|c}
Dataset & Avg. Input (Std) & Avg. Output (Std) \\\hline
Splitwise~\cite{splitwise} & 1155 (1109) & 211 (163) \\
LMSYS-Chat~\cite{zheng2023lmsyschat1m} & 102 (169) & 222 (210) \\
ShareGPT~\cite{sharegpt} & 246 (547) & 322  (244) \\
\end{tabular}
\caption{The average and standard deviation of input and output lengths in the sampled datasets.}
\label{tab:datasets}

\end{table}

\textbf{Datasets.}
Splitwise~\cite{splitwise} is a conversation trace collected from a real production environment at Microsoft, with a total of around 20000 requests.
LMSYS-Chat-1M \cite{zheng2023lmsyschat1m} is a large-scale dataset with 1 million real-world conversations from 25 different LLMs.
ShareGPT~\cite{sharegpt} is a dataset with conversations collected from the ShareGPT API.
We use the full trace from Splitwise and randomly sample $50,000$ requests from ShareGPT and LMSYS-Chat-1M for our evaluation.
Table~\ref{tab:datasets} shows the average input length and output length in tokens for the sampled datasets we use.

\subsection{Throughput}
\label{sec:eval-offline}

We first evaluate throughput to simulate use cases such as benchmarking, information extraction, data wrangling, and form processing~\cite{sheng2023flexgen}.  We sample actual input (prompt tokens) and output (generated tokens) lengths from the datasets and add additional experiments with constant length. We then deploy a \pllm{} instance (with a dense batch size of 2048 in the case of Llama-2-70B, where \pllm{} delivers best performance) 
to measure the \textit{total throughput} defined as the total number of input and output tokens divided by the execution time. We compare the performance of the \pllm{} instance with optimal throughput derived using Equation \ref{eq:optimal-throughput}, where we measure the peak compute capacity using the state-of-the-art GEMM vendor library CUTLASS with a token batch size of $2048$.
 Since the optimal throughput is independent of user query statistics, for all the experiments, the optimal throughput is $1857$ tokens/s/GPU as computed in \S\ref{sec:analysis-optimal}.

\fig{fig:offline-throughput} shows the throughput of the \pllm{} instance compared to the baselines. 
\pllm{} has the highest throughput in all settings and is able to achieve \computeoptimalratio{} of the theoretical optimal throughput in the best case.
For constant lengths, \pllm{} has on average $2.62\times$ higher offline throughput than vLLM, $2.78\times$ higher than DeepSpeed-FastGen and $1.73\times$ higher than TensorRT-LLM. When input and output lengths are drawn from the datasets, \pllm{} has on average $4.18\times$ higher throughput than vLLM, $3.45\times$ higher than DeepSpeed-FastGen and \computesota{} higher than TensorRT-LLM.

\subsection{Latency}
\label{sec:eval-latency}

We evaluate the latency of the \pllm{} instance, for which we model the request arrival interval via an exponential distribution, following prior work \cite{vllm}.  We generate $5$ minutes of request traces from the datasets in Table~\ref{tab:datasets} for various request rates. We then measure \textit{normalized latency} by first calculating the end-to-end request latency divided by \outputlen, then taking the average for all requests. For each framework, we gradually increase the request rate and measure the normalized latency. We select the SLO for average normalized latency as $200ms$ following prior works\cite{distserve}, which is also the typical human reading speed\cite{BRYSBAERT2019104047}.

\fig{fig:online-latency} demonstrates the normalized latency of the \pllm{} instance compared to baselines at various request rates. At lower rates, the \pllm{} instance has comparable but slightly higher
latency, because \pllm{} targets throughput-oriented scenarios and therefore employs a large dense batch size. 

The \pllm{} instance is able to sustain a higher request rate within $200ms$ latency SLO compared to all other baselines across all the datasets. For example, for LMSys-Chat-1M dataset, the \pllm{} instance can achieve \requestsota higher request rate compared to TensorRT-LLM within $200$ms normalized latency constraint.

Moreover, the \pllm{} instance's 99th-percentile latency is only $1.07\times$ of the average latency at near-maximum throughput, as \pllm{} uses a constant dense batch size, helping it perform well even under tail latency SLOs.

\begin{figure}[!t]
  \centering
  \includegraphics[width=\columnwidth]{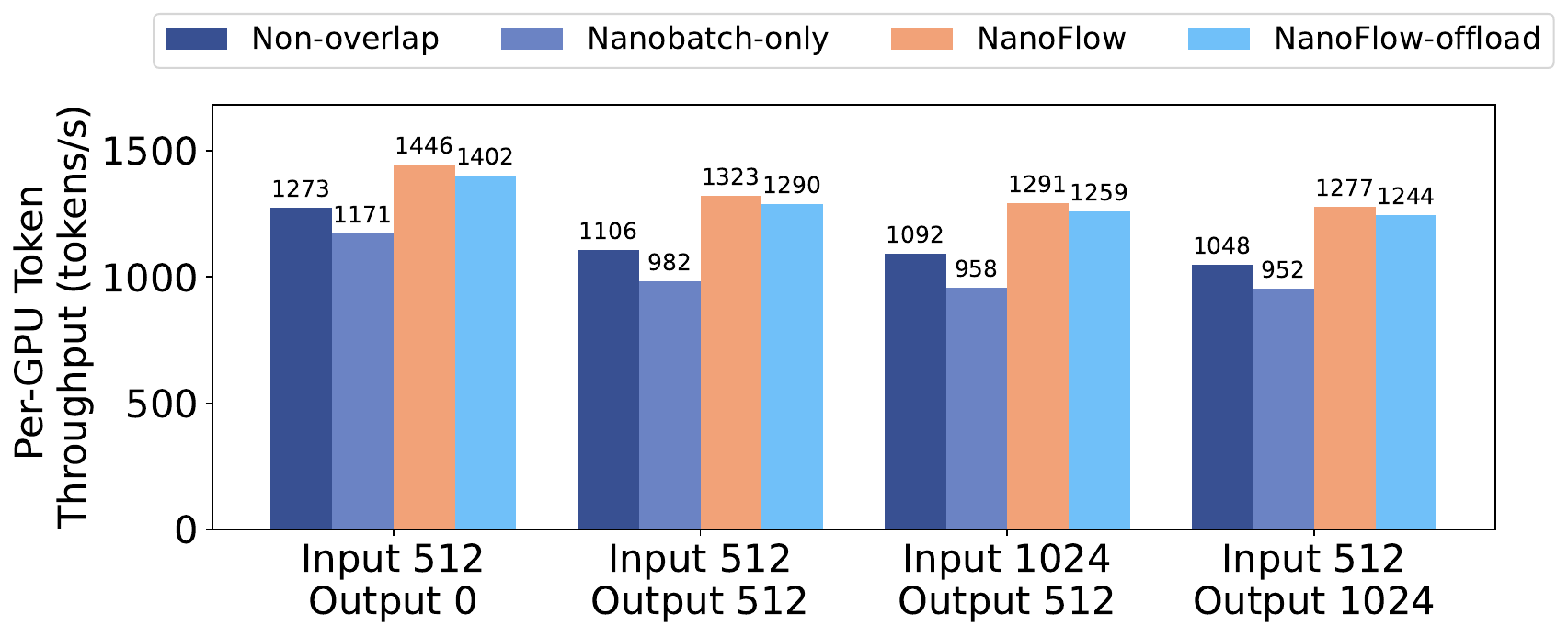}
  \caption{Ablation study results for \pllm{}. \Nano{}ing and overlapping improves \pllm{}'s performance.}
  \label{fig:ablation}
\end{figure}

\subsection{Ablation Study}
\label{sec:eval-ablation}
To showcase the overhead and benefit of the techniques that \pllm{} uses, including \nano{}ing, \nanoop{} overlapping, and \kv{} offloading, we compare the \pllm{} instance with baselines that share the same asynchronous request scheduling and kernel libraries.

We created two baselines: a non-overlapping baseline processing inputs sequentially without \nano{}es, and a \nano{}-only baseline that splits requests into \nano{}es as in \pllm{} but executes them sequentially to assess \nano{}ing overhead. As shown in \fig{fig:ablation}, splitting into \nano{}es alone reduces performance by \nanooverhead.

To estimate the benefit of overlapping network- and memory-bound kernels, we compare \pllm{} and baselines under various prefill-decode ratios, as shown in \fig{fig:ablation}. We use prefill-only workloads (Input 512, Output 0) to demonstrate the benefit of overlapping network-bound and compute-bound kernels.
Additionally, we use decode heavy workloads (Input 512, Output 1024) to evaluate the benefit of overlapping both network- and memory-bound kernels. \pllm{} achieves \benefitnetwork speedup by overlapping network-bound kernels and \benefittotal speedup by overlapping both network- and memory- bound kernels compared with the non-overlapped baseline.

Moreover, we quantify the performance degradation due to offloading in \fig{fig:ablation}. Enabling offloading would slow down the pipeline by $3.0\%$ due to kernel interference caused by \kv{} movement. However, offloading can reduce $3.02x$ compute for multi-round LMSYS-Chat workloads.

\begin{figure*}[!t]
  \centering
  \subfloat[Non-overlap pipeline resource usage]{
  \includegraphics[width=0.49\textwidth]{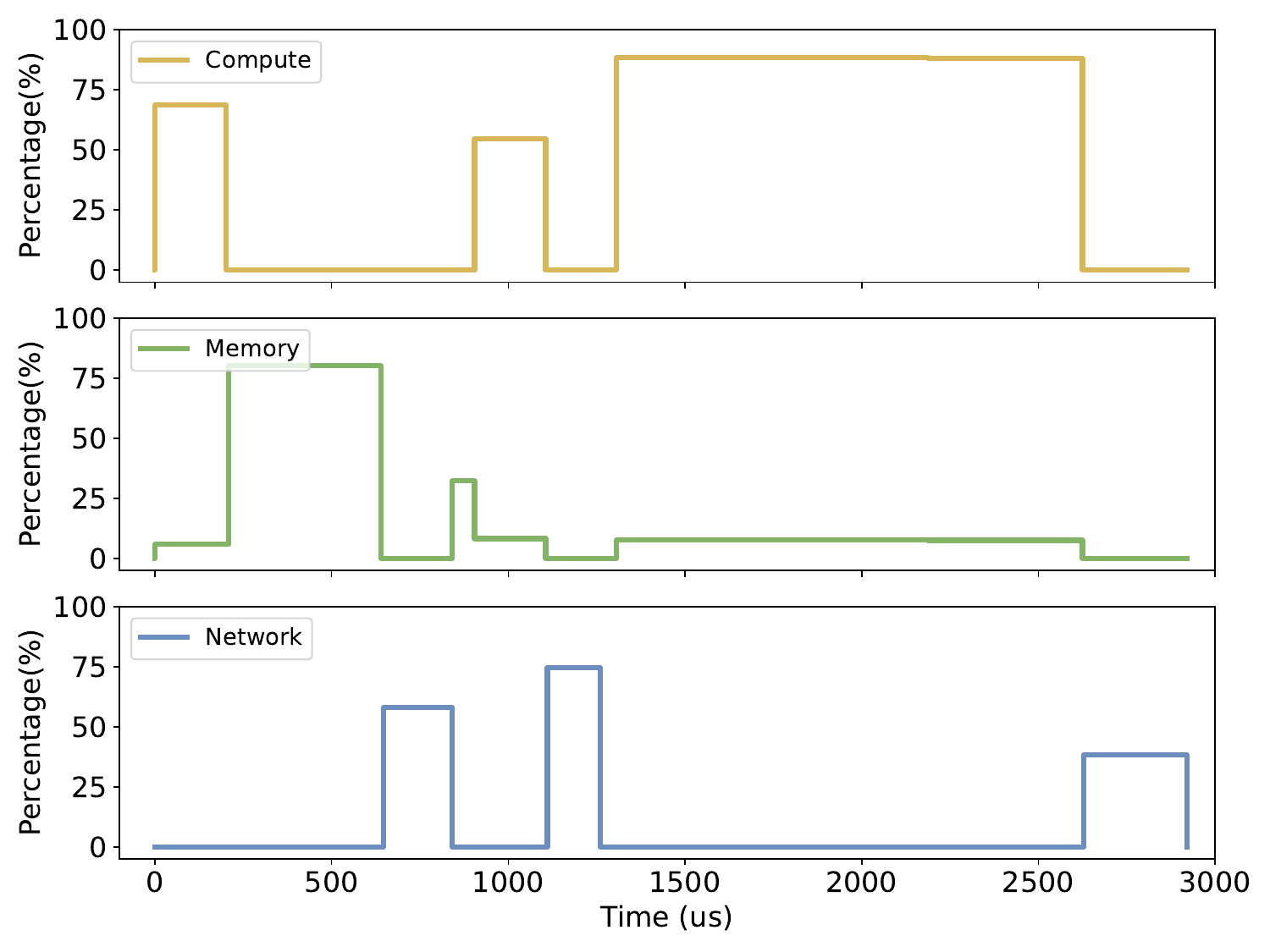}
}
  \subfloat[\pllm{} resource usage]{\includegraphics[width=0.49\textwidth]{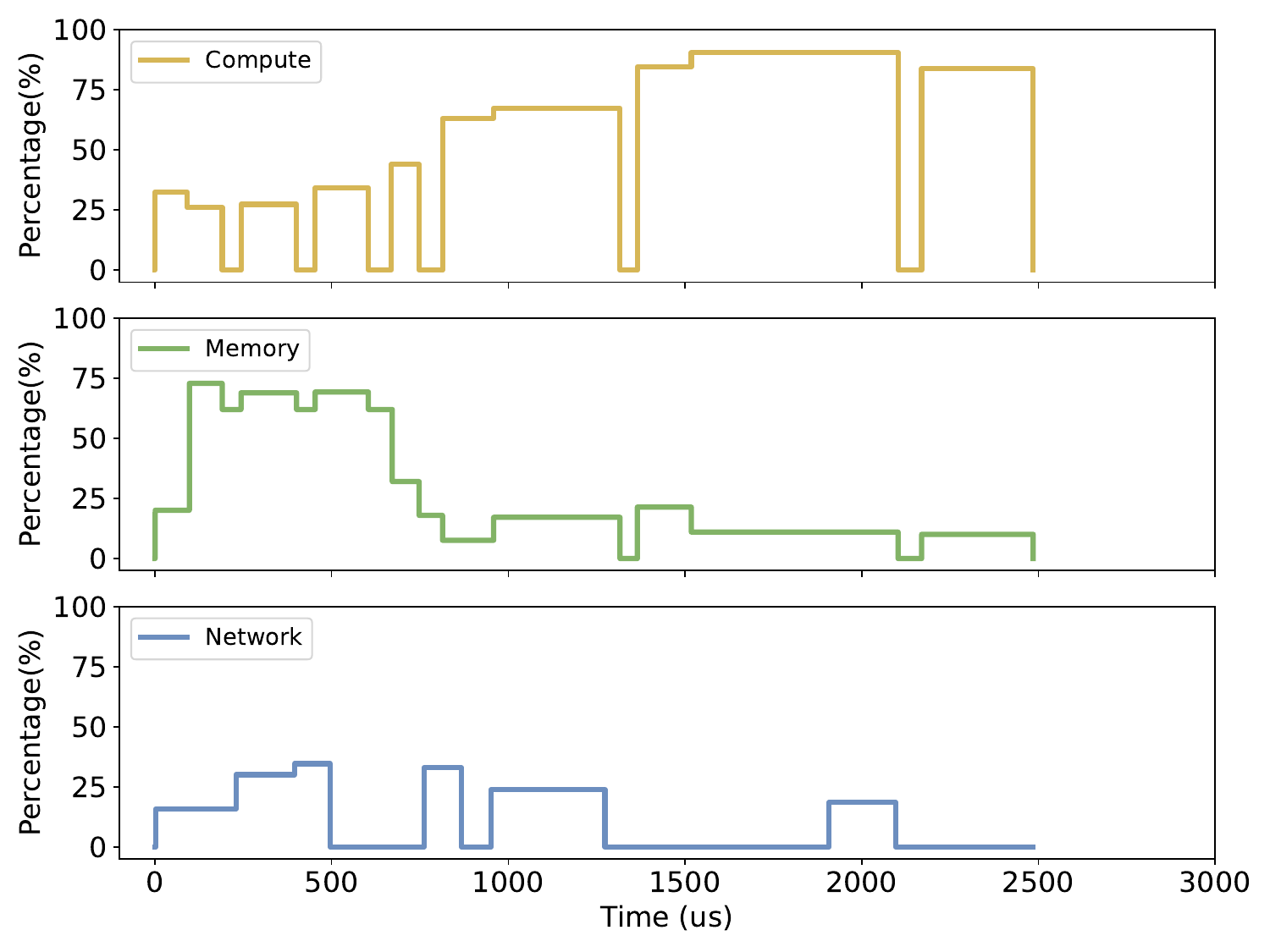}}
  \label{fig:eval-resources-nanoflow}
  \caption{\blue{Non-overlapped pipeline and \pllm{}'s resource usage during inference of a single layer. \pllm{} achieves high compute usage across the whole pipeline by simultaneously also utilizing memory and network bandwidth.}}
  \label{fig:eval-resources}
\end{figure*}

\begin{figure}[t]
  \centering
  \includegraphics[width=\columnwidth]{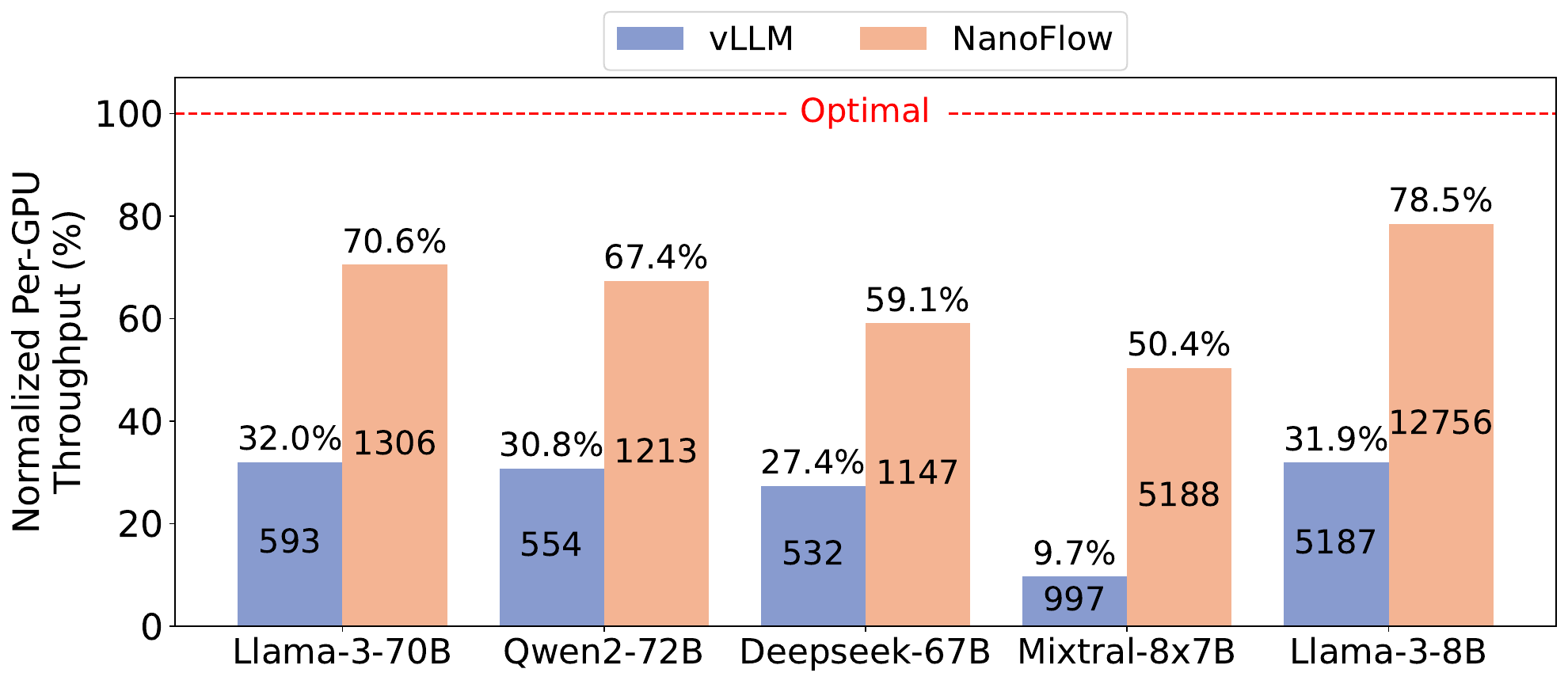}
  \caption{Performance of \pllm{} instances in terms of tokens per second per GPU for other models. Compared with vLLM, \pllm{} significantly increases throughput. \pllm{} achieves up to 78.5\% of optimal throughput.}
  \label{fig:eval-port}
  \vspace{-0.2in}
\end{figure}

\subsection{Resource Usage}
\label{sec:eval-resource}
We demonstrate \pllm{}'s effectiveness in terms of resource utilization in \fig{fig:eval-resources}. 
While the non-overlapping baseline sequentially executes operations, which mostly uses only one resource at a given time, the \pllm{} instance can concurrently utilize multiple resources and achieves \computeoptimalratio average compute utilization. 
Due to kernel interference, \pllm{} provides lower than optimal compute usage.

\subsection{Performance on Other LLMs}
\label{sec:eval-porting}
We evaluate the throughput of \pllm{} instances for other LLMs with a constant length of input $1024$ and output $512$.
All evaluations are done on $8\times$A100 80GB SXM GPUs, except for \lm{}-3-8B, which runs on a single A100 80GB SXM. The results are summarized in \fig{fig:eval-port}. We find that \pllm{} improves throughput to between \portmin and \portmax of optimal throughput for these models, surpassing vLLM.

\section{Related Work}

\textbf{LLM serving optimizations.} Prior works have investigated optimizations for improving the serving throughput at different granularities.
Orca~\cite{orca} proposes request-level continuous batching, which refills the on-the-fly batch to maximize batch size at the granularity of an iteration.
DistServe~\cite{distserve} and Splitwise~\cite{splitwise} explore phase-level scheduling, which disaggregates the prefill and decode phases into different clusters.
DeepSpeed-FastGen~\cite{deepspeed-fastgen} and Sarathi-Serve~\cite{sarathi-serve} propose a chunked prefill policy, which splits a prefill request into multiple smaller chunks and batches decode and prefill requests together to improve overall utilization. However, these works do not perform scheduling at the granularity of operations. \pllm{} exploits \newparall{}, which features fine-grained resource management. Other works tackle the inference inefficiency by optimizing specific operations, including PageAttention~\cite{vllm}, quantization approaches~\cite{atom,qserve,tang2024quest}, etc. \pllm{} can easily adopt new operation implementations by profiling their performances and interference charateristics and integrating them to \auto{}.

\textbf{Operation-level parallelism.}
Some works focus on improving the efficiency of DNN workloads with operation-level optimizations. For example, Rammer~\cite{rammer} explores intra-operation parallelism by remapping the operations into different functional units. Unity~\cite{unity} investigates the combination of parallelism and equivalent algebraic transformation of operations.
However, these works do not consider the different resource demands of operations. 
ASPEN~\cite{aspen} and Welder~\cite{welder} break the boundary between operations by constructing and compiling a tile-level data graph. Since tiles follow sequential dependencies, parallelism is still limited. Moreover, prior works require recompilation from scratch, which requires significant manual work.
Compared with them, \pllm{} creates more overlapping opportunities through \nano{}ing.

\section{Conclusion}
In this work, we analyzed the characteristics of different operations in LLM serving and revealed the compute-bound nature of modern LLM serving workloads. We identified that compute is underutilized inside a single GPU due to the sequential execution of compute-bound, memory-bound, and network-bound operations, leading to sub-optimal throughput.
To address this, we proposed \pllm{}, a novel end-to-end LLM serving system that overlaps operations with heterogeneous resource usages through \newparall{}. 
 \pllm{} automatically constructs a pipeline that overlaps \nanoop{}s, with an auto-search adaptive to various LLM models. Our experiments show that \pllm{} achieves \computesota{} throughput improvements over the state-of-the-art systems.

\bibliographystyle{plain}
\bibliography{_reference}

\begin{thebibliography}{10}

\bibitem{sharegpt}
Sharegpt.
\newblock
  \url{https://huggingface.co/datasets/anon8231489123/ShareGPT_Vicuna_unfiltered},
  2023.

\bibitem{sarathi-serve}
Amey Agrawal, Nitin Kedia, Ashish Panwar, Jayashree Mohan, Nipun Kwatra,
  Bhargav~S Gulavani, Alexey Tumanov, and Ramachandran Ramjee.
\newblock Taming throughput-latency tradeoff in llm inference with
  sarathi-serve.
\newblock {\em arXiv preprint arXiv:2403.02310}, 2024.

\bibitem{sarathi}
Amey Agrawal, Ashish Panwar, Jayashree Mohan, Nipun Kwatra, Bhargav~S Gulavani,
  and Ramachandran Ramjee.
\newblock Sarathi: Efficient llm inference by piggybacking decodes with chunked
  prefills.
\newblock {\em arXiv preprint arXiv:2308.16369}, 2023.

\bibitem{nemo}
Mistral AI.
\newblock Mistral nemo.
\newblock \url{https://mistral.ai/news/mistral-nemo/}, 2024.

\bibitem{ainslie2023gqa}
Joshua Ainslie, James Lee-Thorp, Michiel de~Jong, Yury Zemlyanskiy, Federico
  Lebron, and Sumit Sanghai.
\newblock {GQA}: Training generalized multi-query transformer models from
  multi-head checkpoints.
\newblock In Houda Bouamor, Juan Pino, and Kalika Bali, editors, {\em
  Proceedings of the 2023 Conference on Empirical Methods in Natural Language
  Processing}, pages 4895--4901, Singapore, December 2023. Association for
  Computational Linguistics.

\bibitem{layernorm}
Jimmy~Lei Ba, Jamie~Ryan Kiros, and Geoffrey~E. Hinton.
\newblock Layer normalization, 2016.

\bibitem{brown2020language}
Tom Brown, Benjamin Mann, Nick Ryder, Melanie Subbiah, Jared~D Kaplan, Prafulla
  Dhariwal, Arvind Neelakantan, Pranav Shyam, Girish Sastry, Amanda Askell,
  et~al.
\newblock Language models are few-shot learners.
\newblock {\em Advances in neural information processing systems},
  33:1877--1901, 2020.

\bibitem{BRYSBAERT2019104047}
Marc Brysbaert.
\newblock How many words do we read per minute? a review and meta-analysis of
  reading rate.
\newblock {\em Journal of Memory and Language}, 109:104047, 2019.

\bibitem{qwen}
Alibaba Cloud.
\newblock Alibaba cloud’s qwen2 with enhanced capabilities tops llm
  leaderboard.
\newblock
  \url{https://www.alibabacloud.com/blog/alibaba-cloud’s-qwen2-with-enhanced-capabilities-tops-llm-leaderboard_601268/},
  2024.

\bibitem{silu}
Stefan Elfwing, Eiji Uchibe, and Kenji Doya.
\newblock Sigmoid-weighted linear units for neural network function
  approximation in reinforcement learning, 2017.

\bibitem{nytgpushortage}
Erin Griffith.
\newblock The desperate hunt for the a.i. boom’s most indispensable prize.
\newblock Technical report, The New York Times, 2023.

\bibitem{he2022fastermoe}
Jiaao He, Jidong Zhai, Tiago Antunes, Haojie Wang, Fuwen Luo, Shangfeng Shi,
  and Qin Li.
\newblock Fastermoe: Modeling and optimizing training of large-scale dynamic
  pre-trained models.
\newblock In {\em Proceedings of the 27th ACM SIGPLAN Symposium on Principles
  and Practice of Parallel Programming}, PPoPP '22, page 120–134, New York,
  NY, USA, 2022. Association for Computing Machinery.

\bibitem{deepspeed-fastgen}
Connor Holmes, Masahiro Tanaka, Michael Wyatt, Ammar~Ahmad Awan, Jeff Rasley,
  Samyam Rajbhandari, Reza~Yazdani Aminabadi, Heyang Qin, Arash Bakhtiari, Lev
  Kurilenko, et~al.
\newblock Deepspeed-fastgen: High-throughput text generation for llms via mii
  and deepspeed-inference.
\newblock {\em arXiv preprint arXiv:2401.08671}, 2024.

\bibitem{huang2019gpipe}
Yanping Huang, Youlong Cheng, Ankur Bapna, Orhan Firat, Mia~Xu Chen, Dehao
  Chen, HyoukJoong Lee, Jiquan Ngiam, Quoc~V. Le, Yonghui Wu, and Zhifeng Chen.
\newblock Gpipe: Efficient training of giant neural networks using pipeline
  parallelism, 2019.

\bibitem{jain2025intelligentrouterllmworkloads}
Kunal Jain, Anjaly Parayil, Ankur Mallick, Esha Choukse, Xiaoting Qin, Jue
  Zhang, Íñigo Goiri, Rujia Wang, Chetan Bansal, Victor Rühle, Anoop
  Kulkarni, Steve Kofsky, and Saravan Rajmohan.
\newblock Intelligent router for llm workloads: Improving performance through
  workload-aware load balancing, 2025.

\bibitem{mistral}
Albert~Q. Jiang, Alexandre Sablayrolles, Arthur Mensch, Chris Bamford,
  Devendra~Singh Chaplot, Diego de~las Casas, Florian Bressand, Gianna Lengyel,
  Guillaume Lample, Lucile Saulnier, Lélio~Renard Lavaud, Marie-Anne Lachaux,
  Pierre Stock, Teven~Le Scao, Thibaut Lavril, Thomas Wang, Timothée Lacroix,
  and William~El Sayed.
\newblock Mistral 7b, 2023.

\bibitem{vllm}
Woosuk Kwon, Zhuohan Li, Siyuan Zhuang, Ying Sheng, Lianmin Zheng, Cody~Hao Yu,
  Joseph Gonzalez, Hao Zhang, and Ion Stoica.
\newblock Efficient memory management for large language model serving with
  pagedattention.
\newblock In {\em Proceedings of the 29th Symposium on Operating Systems
  Principles}, pages 611--626, 2023.

\bibitem{alpaserve}
Zhuohan Li, Lianmin Zheng, Yinmin Zhong, Vincent Liu, Ying Sheng, Xin Jin,
  Yanping Huang, Zhifeng Chen, Hao Zhang, Joseph~E Gonzalez, et~al.
\newblock Alpaserve: Statistical multiplexing with model parallelism for deep
  learning serving.
\newblock In {\em 17th USENIX Symposium on Operating Systems Design and
  Implementation (OSDI 23)}, pages 663--679, 2023.

\bibitem{qserve}
Yujun Lin, Haotian Tang, Shang Yang, Zhekai Zhang, Guangxuan Xiao, Chuang Gan,
  and Song Han.
\newblock Qserve: W4a8kv4 quantization and system co-design for efficient llm
  serving, 2024.

\bibitem{rammer}
Lingxiao Ma, Zhiqiang Xie, Zhi Yang, Jilong Xue, Youshan Miao, Wei Cui,
  Wenxiang Hu, Fan Yang, Lintao Zhang, and Lidong Zhou.
\newblock Rammer: Enabling holistic deep learning compiler optimizations with
  {rTasks}.
\newblock In {\em 14th USENIX Symposium on Operating Systems Design and
  Implementation (OSDI 20)}, pages 881--897. USENIX Association, November 2020.

\bibitem{Mehdi_2023}
Yusuf Mehdi.
\newblock Reinventing search with a new ai-powered microsoft bing and edge,
  your copilot for the web, May 2023.

\bibitem{llama3}
Meta.
\newblock Build the future of ai with meta llama 3.
\newblock \url{https://llama.meta.com/llama3/}, 2024.

\bibitem{deepspeedgithub}
Microsoft.
\newblock Deepspeed: Fastgen.
\newblock
  \url{https://github.com/microsoft/DeepSpeed/tree/429bc5c/blogs/deepspeed-fastgen},
  2024.
\newblock Commit ID: 429bc5c, Accessed: 2024-12-09.

\bibitem{dgx}
NVIDIA.
\newblock Nvidia dgx platform.
\newblock \url{https://www.nvidia.com/en-us/data-center/dgx-platform/}, 2024.

\bibitem{nvlink}
NVIDIA.
\newblock Nvlink and nvlink switch.
\newblock \url{https://www.nvidia.com/en-us/data-center/nvlink/}, 2024.

\bibitem{tensorrtllm}
NVIDIA.
\newblock Tensorrt-llm.
\newblock \url{https://github.com/NVIDIA/TensorRT-LLM}, 2024.

\bibitem{tensorrtgithub}
NVIDIA.
\newblock Tensorrt-llm.
\newblock \url{https://github.com/NVIDIA/TensorRT-LLM/tree/5955b8a}, 2024.
\newblock Commit ID: 5955b8a, Accessed: 2024-12-09.

\bibitem{ChatGPT}
OpenAI.
\newblock Introducing chatgpt, 2023.

\bibitem{gpt4}
OpenAI.
\newblock Gpt-4 technical report, 2024.

\bibitem{aspen}
Jongseok Park, Kyungmin Bin, Gibum Park, Sangtae Ha, and Kyunghan Lee.
\newblock Aspen: Breaking operator barriers for efficient parallelization of
  deep neural networks.
\newblock In A.~Oh, T.~Naumann, A.~Globerson, K.~Saenko, M.~Hardt, and
  S.~Levine, editors, {\em Advances in Neural Information Processing Systems},
  volume~36, pages 68625--68638. Curran Associates, Inc., 2023.

\bibitem{seminanalysis}
Dylan Patel and Afzal Ahmad.
\newblock The inference cost of search disruption – large language model cost
  analysis, February 2023.

\bibitem{splitwise}
Pratyush Patel, Esha Choukse, Chaojie Zhang, Íñigo Goiri, Aashaka Shah, Saeed
  Maleki, and Ricardo Bianchini.
\newblock Splitwise: Efficient generative llm inference using phase splitting,
  2023.

\bibitem{patke2025hierarchicalautoscalinglargelanguage}
Archit Patke, Dhemath Reddy, Saurabh Jha, Chandra Narayanaswami, Zbigniew
  Kalbarczyk, and Ravishankar Iyer.
\newblock Hierarchical autoscaling for large language model serving with
  chiron, 2025.

\bibitem{kvcache}
Reiner Pope, Sholto Douglas, Aakanksha Chowdhery, Jacob Devlin, James Bradbury,
  Jonathan Heek, Kefan Xiao, Shivani Agrawal, and Jeff Dean.
\newblock Efficiently scaling transformer inference.
\newblock {\em Proceedings of Machine Learning and Systems}, 5, 2023.

\bibitem{openai_chatgpt_verge}
Emma Roth.
\newblock Chatgpt’s weekly users have doubled in less than a year, 2024.
\newblock Accessed: 2024-12-03.

\bibitem{sheng2023flexgen}
Ying Sheng, Lianmin Zheng, Binhang Yuan, Zhuohan Li, Max Ryabinin, Beidi Chen,
  Percy Liang, Christopher R{\'e}, Ion Stoica, and Ce~Zhang.
\newblock Flexgen: High-throughput generative inference of large language
  models with a single gpu.
\newblock In {\em International Conference on Machine Learning}, pages
  31094--31116. PMLR, 2023.

\bibitem{welder}
Yining Shi, Zhi Yang, Jilong Xue, Lingxiao Ma, Yuqing Xia, Ziming Miao, Yuxiao
  Guo, Fan Yang, and Lidong Zhou.
\newblock Welder: Scheduling deep learning memory access via tile-graph.
\newblock In {\em 17th USENIX Symposium on Operating Systems Design and
  Implementation (OSDI 23)}, pages 701--718, Boston, MA, July 2023. USENIX
  Association.

\bibitem{shoeybi2020megatronlm}
Mohammad Shoeybi, Mostofa Patwary, Raul Puri, Patrick LeGresley, Jared Casper,
  and Bryan Catanzaro.
\newblock Megatron-lm: Training multi-billion parameter language models using
  model parallelism, 2020.

\bibitem{megatron-lm}
Mohammad Shoeybi, Mostofa Patwary, Raul Puri, Patrick LeGresley, Jared Casper,
  and Bryan Catanzaro.
\newblock Megatron-lm: Training multi-billion parameter language models using
  model parallelism, 2020.

\bibitem{demandsageChatGPTStatistics}
Shubham Singh.
\newblock {C}hat{G}{P}{T} {S}tatistics ({A}{U}{G} 2024) - {U}sers {G}rowth
  {D}ata --- demandsage.com.
\newblock \url{https://www.demandsage.com/chatgpt-statistics/}, 2024.
\newblock [Accessed 14-08-2024].

\bibitem{Spataro_2023}
Jared Spataro.
\newblock Introducing microsoft 365 copilot – your copilot for work, May
  2023.

\bibitem{cpuoverhead}
Vikranth Srivatsa, Dongming Li, Yiying Zhang, and Reyna Abhyankar.
\newblock {M}{L}{S}ys @ {W}uk{L}ab - {C}an {S}cheduling {O}verhead {D}ominate
  {L}{L}{M} {I}nference {P}erformance? {A} {S}tudy of {C}{P}{U} {S}cheduling
  {O}verhead on {T}wo {P}opular {L}{L}{M} {I}nference {S}ystems ---
  mlsys.wuklab.io.
\newblock \url{https://mlsys.wuklab.io/posts/scheduling_overhead/}, 2024.
\newblock [Accessed 25-10-2024].

\bibitem{orion}
Foteini Strati, Xianzhe Ma, and Ana Klimovic.
\newblock Orion: Interference-aware, fine-grained gpu sharing for ml
  applications.
\newblock In {\em Proceedings of the Nineteenth European Conference on Computer
  Systems}, EuroSys '24, page 1075–1092, New York, NY, USA, 2024. Association
  for Computing Machinery.

\bibitem{tang2024quest}
Jiaming Tang, Yilong Zhao, Kan Zhu, Guangxuan Xiao, Baris Kasikci, and Song
  Han.
\newblock Quest: Query-aware sparsity for efficient long-context llm inference,
  2024.

\bibitem{theaibrixteam2025aibrixscalablecosteffectivelarge}
The~AIBrix Team, Jiaxin Shan, Varun Gupta, Le~Xu, Haiyang Shi, Jingyuan Zhang,
  Ning Wang, Linhui Xu, Rong Kang, Tongping Liu, Yifei Zhang, Yiqing Zhu,
  Shuowei Jin, Gangmuk Lim, Binbin Chen, Zuzhi Chen, Xiao Liu, Xin Chen, Kante
  Yin, Chak-Pong Chung, Chenyu Jiang, Yicheng Lu, Jianjun Chen, Caixue Lin,
  Wu~Xiang, Rui Shi, and Liguang Xie.
\newblock Aibrix: Towards scalable, cost-effective large language model
  inference infrastructure, 2025.

\bibitem{Thakkar_CUTLASS_2023}
Vijay Thakkar, Pradeep Ramani, Cris Cecka, Aniket Shivam, Honghao Lu, Ethan
  Yan, Jack Kosaian, Mark Hoemmen, Haicheng Wu, Andrew Kerr, Matt Nicely, Duane
  Merrill, Dustyn Blasig, Fengqi Qiao, Piotr Majcher, Paul Springer, Markus
  Hohnerbach, Jin Wang, and Manish Gupta.
\newblock {CUTLASS}, January 2023.

\bibitem{gpushortage}
CHENG TING-FANG.
\newblock Tsmc sees ai chip output constraints lasting 1.5 years.
\newblock Technical report, Nikkei Asia, 2023.

\bibitem{llama}
Hugo Touvron, Thibaut Lavril, Gautier Izacard, Xavier Martinet, Marie-Anne
  Lachaux, Timothée Lacroix, Baptiste Rozière, Naman Goyal, Eric Hambro,
  Faisal Azhar, Aurelien Rodriguez, Armand Joulin, Edouard Grave, and Guillaume
  Lample.
\newblock Llama: Open and efficient foundation language models, 2023.

\bibitem{llama2}
Hugo Touvron, Louis Martin, Kevin Stone, Peter Albert, Amjad Almahairi, Yasmine
  Babaei, Nikolay Bashlykov, Soumya Batra, Prajjwal Bhargava, Shruti Bhosale,
  et~al.
\newblock Llama 2: Open foundation and fine-tuned chat models.
\newblock {\em arXiv preprint arXiv:2307.09288}, 2023.

\bibitem{unity}
Colin Unger, Zhihao Jia, Wei Wu, Sina Lin, Mandeep Baines, Carlos
  Efrain~Quintero Narvaez, Vinay Ramakrishnaiah, Nirmal Prajapati, Pat
  McCormick, Jamaludin Mohd-Yusof, Xi~Luo, Dheevatsa Mudigere, Jongsoo Park,
  Misha Smelyanskiy, and Alex Aiken.
\newblock Unity: Accelerating {DNN} training through joint optimization of
  algebraic transformations and parallelization.
\newblock In {\em 16th USENIX Symposium on Operating Systems Design and
  Implementation (OSDI 22)}, pages 267--284, Carlsbad, CA, July 2022. USENIX
  Association.

\bibitem{attention}
Ashish Vaswani, Noam Shazeer, Niki Parmar, Jakob Uszkoreit, Llion Jones,
  Aidan~N. Gomez, \L{}ukasz Kaiser, and Illia Polosukhin.
\newblock Attention is all you need.
\newblock In {\em Proceedings of the 31st International Conference on Neural
  Information Processing Systems}, NIPS'17, page 6000–6010, Red Hook, NY,
  USA, 2017. Curran Associates Inc.

\bibitem{vllmgithub}
vllm project.
\newblock vllm.
\newblock \url{https://github.com/vllm-project/vllm/tree/38c4b7e}, 2024.
\newblock Commit ID: 38c4b7e, Accessed: 2024-12-09.

\bibitem{orca}
Gyeong-In Yu, Joo~Seong Jeong, Geon-Woo Kim, Soojeong Kim, and Byung-Gon Chun.
\newblock Orca: A distributed serving system for $\{$Transformer-Based$\}$
  generative models.
\newblock In {\em 16th USENIX Symposium on Operating Systems Design and
  Implementation (OSDI 22)}, pages 521--538, 2022.

\bibitem{zhao2024forestcollefficientcollectivecommunications}
Liangyu Zhao, Saeed Maleki, Aashaka Shah, Ziyue Yang, Hossein Pourreza, and
  Arvind Krishnamurthy.
\newblock Forestcoll: Efficient collective communications on heterogeneous
  network fabrics, 2024.

\bibitem{atom}
Yilong Zhao, Chien-Yu Lin, Kan Zhu, Zihao Ye, Lequn Chen, Size Zheng, Luis
  Ceze, Arvind Krishnamurthy, Tianqi Chen, and Baris Kasikci.
\newblock Atom: Low-bit quantization for efficient and accurate llm serving.
\newblock {\em arXiv preprint arXiv:2310.19102}, 2023.

\bibitem{zheng2023lmsyschat1m}
Lianmin Zheng, Wei-Lin Chiang, Ying Sheng, Tianle Li, Siyuan Zhuang, Zhanghao
  Wu, Yonghao Zhuang, Zhuohan Li, Zi~Lin, Eric.~P Xing, Joseph~E. Gonzalez, Ion
  Stoica, and Hao Zhang.
\newblock Lmsys-chat-1m: A large-scale real-world llm conversation dataset,
  2023.

\bibitem{alpa}
Lianmin Zheng, Zhuohan Li, Hao Zhang, Yonghao Zhuang, Zhifeng Chen, Yanping
  Huang, Yida Wang, Yuanzhong Xu, Danyang Zhuo, Eric~P Xing, et~al.
\newblock Alpa: Automating inter-and $\{$Intra-Operator$\}$ parallelism for
  distributed deep learning.
\newblock In {\em 16th USENIX Symposium on Operating Systems Design and
  Implementation (OSDI 22)}, pages 559--578, 2022.

\bibitem{zheng2024sglangefficientexecutionstructured}
Lianmin Zheng, Liangsheng Yin, Zhiqiang Xie, Chuyue Sun, Jeff Huang, Cody~Hao
  Yu, Shiyi Cao, Christos Kozyrakis, Ion Stoica, Joseph~E. Gonzalez, Clark
  Barrett, and Ying Sheng.
\newblock Sglang: Efficient execution of structured language model programs,
  2024.

\bibitem{distserve}
Yinmin Zhong, Shengyu Liu, Junda Chen, Jianbo Hu, Yibo Zhu, Xuanzhe Liu, Xin
  Jin, and Hao Zhang.
\newblock Distserve: Disaggregating prefill and decoding for goodput-optimized
  large language model serving.
\newblock {\em arXiv preprint arXiv:2401.09670}, 2024.

\end{thebibliography}

\end{document}